\documentclass[secnumarabic,aps,prd,floatfix,nofootinbib,showkeys,twocolumn,showpacs,10pt,superscriptaddress]
{revtex4-1}

\usepackage{times}
\usepackage{amsfonts}
\usepackage{amssymb}
\usepackage{amsmath}
\usepackage{natbib}

\usepackage{graphicx}
\usepackage[tight,footnotesize]{subfigure}
\usepackage{enumitem}

\usepackage{rotating}

\usepackage{xcolor}
\usepackage{color}

\usepackage{setspace}

\usepackage[bookmarks=true,colorlinks=true,citecolor=blue,linkcolor=blue,urlcolor=magenta]{hyperref}

\def\aap{A\&A}
\def\aj{AJ}
\def\apj{ApJ}

\def\apjs{ApJS}

\def\jcap{JCAP}
\def\mnras{MNRAS}

\def\nar{New Astronomy Reviews}

\def\physrep{Physics Reports}

\ifpdf
\graphicspath{
	{figures/}
	{figures/Mt_vs_Rom/}
	{figures/NS_WD_BRANCHES_evolBold/gray_diffColorPalette/}
	{figures/NS_WD_BRANCHES_evolBold/nogray_diffColorPalette/}
	{figures/pressure_vs_mass/}			
	{figures/pressure_vs_mass/dmtype2_y01/eps/}			
	{figures/pressure_vs_mass/dmtype6_y103/eps/}	
	{figures/pressure_vs_StabCriteria_new/}
	{figures/pressure_vs_StabCriteria_new/eps/}
          {figures/radNM_vs_massTotal_pressure_masses_y01/}
          {figures/radNM_vs_massTotal_pressure_masses_y01/eps/}
          {figures/radNM_vs_massTotal_pressure_masses_y103/}
          {figures/radNM_vs_massTotal_pressure_masses_y103/eps/}
}
\else
\graphicspath{
	{figures/}
	{figures/Mt_vs_Rom/}
	{figures/NS_WD_BRANCHES_evolBold/gray_diffColorPalette/}
	{figures/NS_WD_BRANCHES_evolBold/nogray_diffColorPalette/}
	{figures/pressure_vs_mass/}			
	{figures/pressure_vs_mass/dmtype2_y01/eps/}			
	{figures/pressure_vs_mass/dmtype6_y103/eps/}	
	{figures/pressure_vs_StabCriteria_new/}
	{figures/pressure_vs_StabCriteria_new/eps/}
          {figures/radNM_vs_massTotal_pressure_masses_y01/}
          {figures/radNM_vs_massTotal_pressure_masses_y01/eps/}
          {figures/radNM_vs_massTotal_pressure_masses_y103/}
          {figures/radNM_vs_massTotal_pressure_masses_y103/eps/}
}
\fi

\begin{document}

\title{Dark compact objects: an extensive overview}

\author{Maksym ~\surname{Deliyergiyev}}
\affiliation{Institute of Modern Physics, Chinese Academy of Sciences, Department of High Energy Nuclear Physics, Post Office Box31, Lanzhou 730000, Peoples Republic of China}
\affiliation{Institute of Physics, Jan Kochanowski University, PL-25406 Kielce, Poland}
\email{maksym.deliyergiyev@ujk.edu.pl}
\author{Antonino ~\surname{Del Popolo}}
\affiliation{Dipartimento di Fisica e Astronomia, University Of Catania, Viale Andrea Doria 6, 95125, Catania, Italy}
\affiliation{INFN Sezione di Catania, Via S. Sofia 64, I-95123 Catania, Italy}
\email{adelpopolo@oact.inaf.it}
\affiliation{Institute of Modern Physics, Chinese Academy of Sciences, Post Office Box31, Lanzhou 730000, Peoples Republic of China}
\author{Laura ~\surname{Tolos}}
\affiliation{Institut f\"{u}r Theoretische Physik, Goethe Universit\"{a}t Frankfurt, Max-von-Laue-Stra\ss{}e 1, 60438 Frankfurt, Germany}
\affiliation{Frankfurt Institute for Advanced Studies, Goethe Universit\"{a}t Frankfurt, Ruth-Moufang-Str.1, 60438 Frankfurt am Main, Germany}
\affiliation{Institute of Space Sciences (ICE, CSIC), Campus UAB, Carrer de Can Magrans, 08193, Barcelona, Spain}
\affiliation{Institut d’Estudis Espacials de Catalunya (IEEC), 08034 Barcelona, Spain}
\email{tolos@th.physik.uni-frankfurt.de}
\author{Morgan~\surname{Le~Delliou}}
\affiliation{Institute of Theoretical Physics, Physics Department, Lanzhou University, 
No.222, South Tianshui Road, Lanzhou, Gansu 730000, P R China}
\affiliation{Instituto de Astrof\'isica e Ci\^encias do Espa\c co, Universidade de Lisboa, Faculdade de Ci\^encias, 
Ed. C8, Campo Grande, 1769-016 Lisboa, Portugal}
\email[Corresponding author: ]{(delliou@lzu.edu.cn,)delliou@ift.unesp.br}
\author{Xiguo~\surname{Lee}}
\affiliation{Institute of Modern Physics, Chinese Academy of Sciences, Post Office Box31, Lanzhou 730000, Peoples Republic of China}
\email{xgl@impcas.ac.cn}
\author{Fiorella~\surname{Burgio}}
\affiliation{INFN Sezione di Catania, Via S. Sofia 64, I-95123 Catania, Italy}
\email{fiorella.burgio@ct.infn.it}

\label{firstpage}

\date{\today}

\begin{abstract}
  We study the structure of compact objects that contain non-self annihilating, self-interacting dark matter admixed with ordinary matter made of neutron star and white dwarf materials.  We extend {
    the previous work Phys.\ Rev.\ D {\bf 92}  123002 (2015)} on these dark compact objects by analyzing the effect of weak and strongly interacting dark matter with particle masses in the range of 1-500 GeV, {
    so as to set some constraints in the strength of the interaction and the mass of the dark matter particle.} We find that the total mass of the compact objects increases with decreasing dark matter particle mass.  In the strong interacting case and for dark matter particle masses in the range 1-10 GeV, the total mass of the compact objects largely exceeds the $2M_{\odot}$  constraint for neutron star masses and the nominal 1$M_{\odot}$ for white dwarfs, while {
    for larger dark matter particle masses or in the weakly interacting case} the compact objects show masses in agreement or smaller than these constraints, thus hinting at
  the exclusion of strongly self-interacting dark matter of masses 1-10 GeV in the interior of these compact objects. Moreover, we observe that the smaller the dark matter particle mass, the larger the quantity of dark matter captured is, putting constraints on the dark matter mass trapped in the compact objects {
    so as to fullfill $\simeq 2 M_{\odot}$ observations.}
Finally, the inhomogeneity of distribution of dark matter in the Galaxy implies a mass dependence of compact objects from the environment which can be used to put constraints on the characteristics of the Galaxy halo DM profile and on particle mass. In view of the these results, we discuss the formation of the dark  compact objects in an homogeneous and non-homogeneous dark matter environment.
%
\end{abstract}


\maketitle

\section{Introduction}

Astrophysical and cosmological observations strongly indicate that the mass content of our universe is dominated by non-baryonic mass/energy \citep{Bertone2005,DelPopolo2014}. According to \citep{Planck2016}, the Universe is composed of $4.9\%$ of baryonic matter, $26.4\%$ of invisible form of matter whose existence is inferred from its gravitational effects, dubbed dark matter (DM). A further component, dubbed dark energy (DE), which is hypothesized to permeate all space, and whose existence is related to the accelerated expansion of the Universe \citep{Riess1998,Perlmutter1999}, constitutes $68.7\%$ of the total mass.  

In the $\Lambda$CDM model, a parameterization of the big-bang cosmology with six parameters, the DE is associated with the cosmological constant $\Lambda$, and the material components of the Universe are the ones indicated previously. The quoted $\Lambda$CDM paradigm, describes correctly many of the observations\citep{Komatsu2011,Planck2014,DelPopolo2013,DelPopolo2007,DelPopolo2014}, but has some drawbacks. On large scale CMB shows some anomalies, such as that the Planck 2015 data are in tension with the CFHTLenS  weak lensing \citep{Raveri2016}, and $\sigma_8$ \citep{Macaulay2013}. There is also another tension with the value of the Hubble parameter measured by SNIa\footnote{The $\Lambda$CDM paradigm has some other drawbacks, as the cosmological constant problem \citep{Weinberg1989,Astashenok2012}, the unknown nature of DE \citep{DelPopolo2013a,DelPopolo2013b,DelPopolo2013c} and the so called "small scale problems" \citep{DelPopolo2017}.}. Another big issue of the paradigm is the nature of DM. A "zoo" of candidates have been proposed, with masses in the range  of $10^{-33}$ GeV (Fuzzy DM) to $10^{15}$ GeV (Wimpzillas). In that large "zoo", WIMPs, Axions, and sterile neutrinos, have received a peculiar attention \citep{Bertone2005,DelPopolo2014}. 

Several different attempts to detect DM have been made. Direct detection attempts to detect DM particles through their elastic scattering with nuclei (normal matter recoiling from DM collisions). The recoil is not measured directly {
  but  through crystal or liquid scintillation} (DAMA/LIBRA \citep{Bernabei2008}, KIMS\citep{2014KIMS}, CRESST-II \citep{2016CRESST-II}, ZEPLIN \citep{2005ZEPLIN}), phonons generation (CRESST-I \citep{1999CRESST}), ionization (CDMS \citep{2009CDMS}, superCDMS \citep{2018SuperCDMS}, XENON100 \citep{2012Xenon100}, XENON1T \citep{2018Xenon1T}, LUX \citep{2013LUX}), axion cavities (ADMX \citep{2018ADMX}), and several others.  The indirect searches aimt a detecting the products of WIMP annihilations (e.g., gamma rays, neutrinos, positrons, electrons, and antiprotons). {
  The FERMI-LAT, DAMPE \citep{Dampe2017}, CALET \citep{Calet2017}, HAWC \citep{2015HAWC}, HESS \citep{2016HESS}, VERITAS \citep{VERITAS}, MAGIC \citep{MAGIC} or the planned CTA \citep{CTA} are working in the photons channel, while IceCube \cite{IceCube} in neutrinos and FERMI-LAT \cite{FERMI-LAT} in antimatter. 
} Concerning particle accelerators, DM (WIMPs) production together with jets (other particles) should give rise to a large amount of energy. In 2015, in LHC it was observed a 750 GeV resonance in the di-photon final state, later disproved \citep{Sato2016}.  However, all these experiments based on direct or indirect detection \citep{Bertone2005,DelPopolo2014,Klasen2015}, or detection in particle accelerators  \citep{Bednyakov2015} have produced no evidence of the existence of DM particles, apart some claims (e.g. DAMA/LIBRA) not confirmed by other experiments. 

In this context, different avenues for testing the possible effects of DM are welcome. For this purpose, compact objects (COs), such as white dwarfs (WDs) and neutron stars (NSs), present the advantage of extreme densities, increasing the probability of the interaction of DM with nucleons and the DM capture
\citep{Goldman1989, Kouvaris2008}.

The DM (WIMPs) accumulation\footnote{The DM content of a CO depends on a) its formation process and b) the accumulation through capture in the CO's lifetime \citep{Ciarcelluti2010}.} in COs modifies their structure. If the accumulated DM is larger than a critical value \citep{Kouvaris2013}, DM can become self-gravitating forming a mini black-hole in the collapse. This could be used to infer constraints on the mass and cross section of DM (WIMPs) \citep{Bertone2008}. If DM is constituted by WIMPs trapped in the COs, its annihilation produces a heating of the star, and an increase of the surface temperature and luminosity. In the limit case of COs located close to the galactic center, the temperature can reach $10^6$ K, and luminosities $10^{-2} L_{\odot}$ \citep{deLavallaz2010}. The quoted changes are difficult to observe, especially for objects located close to the galactic center \citep{Sandin2009}. 

An appealing alternative to WIMPs is the asymmetric dark matter (ADM) model, based on the idea that the present DM abundance has a similar origin of visible matter \citep{Petraki2013}. A peculiar case of ADM is mirror matter. WIMPs are supersymmetric particles, based on the assumption of a symmetry between bosons and fermions. If one assumes that instead of the quoted symmetry, one has that nature is parity symmetric, we have a different form of DM, mirror matter \cite{Chacko:2005pe, Kobzarev:1966qya}. The main motivation for the existence of mirror matter is that of restoring parity symmetry in nature laws\footnote{Weak interaction are not parity symmetric.}. An interesting feature of this kind of DM is that it is consistent with DAMA experiment \citep{Bernabei2008,Foot2008}. Since ADM is non-annihilating it can accumulate and thermalize in a small radius, producing changes in mass and radius of the stars, with the possibility of forming extraordinary compact NSs. {
Comparing the mass-radius ($M-R$) relation predicted by stars models with ordinary matter and with ordinary matter admixed with DM in NSs, it is possible to extract information on DM and the equation of state (EoS) of the NSs \citep{Ciarcelluti2010}.} These extra compact NSs, having a DM core, could explain the discrepancy between the structure of (e.g.)  4U $1608-52$ and the $M-R$ relationship coming from nuclear matter models \citep{Ciarcelluti2010}.

Moreover, the NSs behaviour is dictated by their EoS, constrained at normal nuclear saturation density \citep{Tews2013}, but not at densities larger than normal. At those densities the properties of matter are unknown. {
This implies that fundamental quantities like the mass cannot be known with certainty. In order to explain NSs with high masses, like the binary millisecond pulsar PSR J1614-2230 of  $M=1.97 \pm 0.04 M_{\odot}$ \citep{Demorest:2010bx} and the PSR J0348+0432 of $M=2.01\pm 0.04 M_{\odot}$ \citep{Antoniadis:2013pzd}, one needs a stiff EoS. Then, whereas most of the phenomenological models for EoS are able to reproduce the $2 M_{\odot}$ observations, these observations are in tension with microscopical models that implement some possible exotic components of the EoS (e.g., quarks, mesons, hyperons) that soften the EoS. }

In recent years, it has been realized that the presence of ADM in NSs plays a similar role to that of exotic states \citep{Sandin2009}. In Ref.~\citep{Sandin2009} the authors considered the NSs composed of ``normal" matter admixed with mirror matter coupled only through gravity, whereas in Ref.~\citep{Leung2011} a general relativistic two fluid approach to study admixture of nuclear matter and degenerate DM was used. In both papers it was found that the presence of DM gives rise to a different $M-R$ relationship, characterized by smaller mass and radii when the increasing ratio of the DM and normal matter increased. In Ref.~\citep{Mukhopadhyay2016} quark matter admixed DM was studied with a similar formalism, finding, among other results, a total mass of 1.95 $M_{\odot}$, close to the 2$M_{\odot}$ observations.  Li et al.  \citep{Li2012} showed that a NS  with ADM has a $M-R$ relationship similar to that obtained by \citep{Sandin2009}, and that small values of the mass of the DM particles produce an increase in the final stellar mass reaching values even larger than $2 M_{\odot}$. 

{
  In the previous work of Ref.~\citep{Tolos2015}, the Tolman-Oppenheimer-Volkoff (TOV) equations were solved for NS and WD material admixed with ADM and particle mass equal to 100 GeV. It was found that in the case of weak self-interacting, non-annihilating DM, the TOV's solutions can give COs with Earth-like masses and radii a few km to a few hundreds km, while in the case of strong self-interactions, Jupiter-like COs with radii of a few hundreds km were obtained.
The maximum DM content that NSs with $2M_{\odot}$ and WDs of the nominal mass of $1M_{\odot}$ can sustain was also analyzed.
   
In the present paper we aim 
to extend this previous work by considering weak and strongly interacting DM with particle masses in the range of 1-500 GeV.  We find that the total mass of the COs increases with decreasing DM particle mass, thus hinting 
at excluding  strongly self-interacting DM of masses 1-10 GeV in the interior of these COs. Moreover, we obtain that the smaller
the mass of the DM particle%
, the more DM is captured in the COs, {
  setting constraints on DM capture given by $2 M_{\odot}$ observations}. This finding has to be tested by analyzing the quantity of DM captured during the formation of these COs inside a DM environment, thus again helping to constrain the DM mass particle and its self-interaction. 

The paper is organized as follows. In Sec.~\ref{sec:Implementation} we describe the theoretical model, whereas in Sec.~\ref{sec:Results} we describe the results. Sec.~\ref{sec:Matter in the COs} contains the description of the capture mechanism which can accrete the mass predicted by the TOV equations solution, giving some constraints on the DM captured.
        Finally Sec.~\ref{sec:Conclusions} is devoted to the discussion and conclusions.}

\section{Theoretical Model}
\label{sec:Implementation}

In the following we solve the TOV equations for an admixture of ADM and ordinary matter (OM) coupled only by gravity. The aim is to understand what kind of COs can be formed, since the spherical configurations obtained by solving the TOV equations may yield to certain stable configurations with unusual masses and radii.  Therefore, we start from the dimensionless TOV equations \cite{Narain:2006kx}, following \citep{Tolos2015} notation,
\begin{eqnarray}
&&\frac{dp'_{OM}}{dr}=-(p'_{OM}+\rho'_{OM})\frac{d \nu}{dr}, \nonumber \\
&&\frac{dm_{OM}}{dr}=4 \pi r^2 \rho'_{OM}, \nonumber \\
&&\frac{dp'_{DM}}{dr}=-(p'_{DM}+\rho'_{DM}) \frac{d \nu}{dr}, \nonumber \\
&&\frac{dm_{DM}}{dr}=4 \pi r^2 \rho'_{DM}, \nonumber \\
&&\frac{d \nu}{dr}=\frac{(m_{OM}+m_{DM}) + 4 \pi r^3(p'_{OM}+p'_{DM})}{r(r-2(m_{OM}+m_{DM}))}, 
\label{eq:TOV_with_DM}
\end{eqnarray}
the quantity $p'=P/m_f^4$, and $\rho'=\rho/m_f^4$, are respectively the dimensionless pressure and energy density, being 
$m_f$ the fermion mass (i.e., DM particle mass, and neutron mass). 
Indicated with $M_p$ the Planck mass \cite{Narain:2006kx}, each one of the two species can give rise to an astrophysical object with radius, 
$R=(M_p/m_f^2) \, r$ and mass $M= (M_p^3/m_f^2) \,  m$.

{
  For OM we use the same EoSs as described in \citep{Tolos2015}, whereas the interacting Fermi gas EoS for DM is taken from Ref.~\cite{Narain:2006kx}. The DM particles are non-self annihilating \cite{Nussinov:1985xr,Kaplan:1991ah,Hooper:2004dc,Kribs:2009fy,Kouvaris:2015rea}, and self-interacting \cite{Spergel:1999mh}. Differently from \citep{Tolos2015}, where
the only DM mass considered was 100 GeV, in this paper we take into account DM particle masses equal to 1, 5, 10, 50, 100, 200, 500 GeV.  } We study the case of weakly interacting, and strong interacting DM. The interaction strength is expressed in terms of the ratio of the fermion mass $m_f$, and scale of interaction $m_I$, $y=m_f/m_I$. The two values of $y$ are $y=10^{-1}$ for weakly interacting DM ($m_I \sim 100~{\rm MeV}$), and $y=10^3$ for strongly interacting DM ($m_I \sim 300~{\rm GeV}$). The ratio between the DM pressure and that of the ordinary matter, $p_{DM}/p_{OM}$, is assumed to be in the range of $10^{-5}$ to  $10^{5}$, a range larger than the one considered in Ref.~\citep{Tolos2015}.
\begin{figure*}[]
\begin{center}
\includegraphics[scale=0.36]{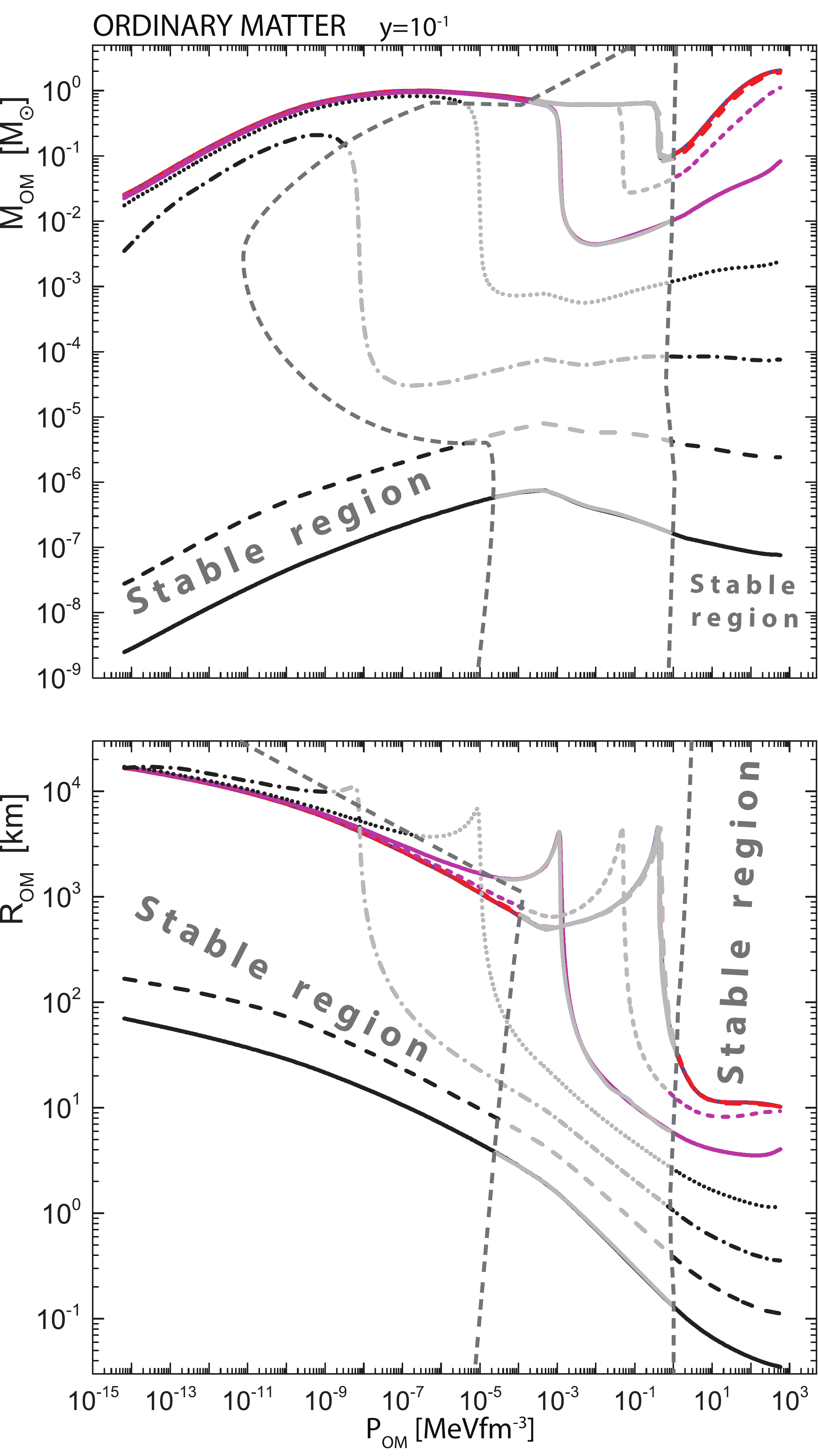}
\includegraphics[scale=0.36]{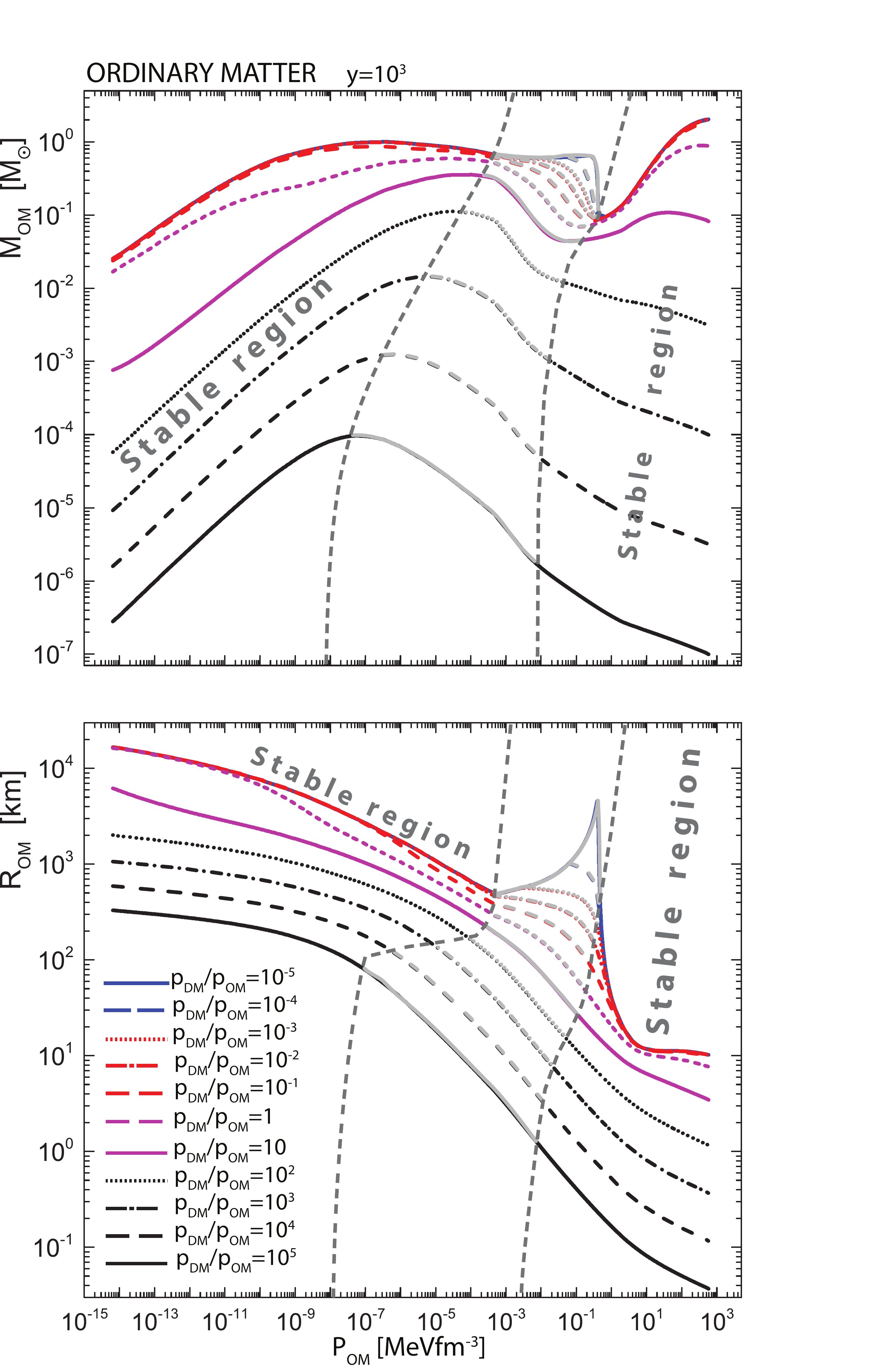}
\caption{
Mass and radius of OM content of the astrophysical objects as a function of the central pressure of OM for the DM interaction strength parameters $y=10^{-1}$(left panels) and $y=10^{3}$(right panels).
The solutions go through both stable and unstable regions. The separations are delimited by the roughly vertical dashed curves.
The coloured lines indicate solutions for $M_{OM}$ and $R_{OM}$ over OM central pressures for a set of dimensionless ratio, $p_{DM}/p_{OM}$, running from $10^{-5}$ to $10^{5}$. {
  Those lines turn grey in the unstable regions. }
Note that for the case $y=10^{-1}$, lines of $p_{DM}/p_{OM} = 10^{-1}$  to  $p_{DM}/p_{OM} = 10^{-5}$ for the $M_{OM}$ and $R_{OM}$ distributions lie on top of each other.
}
\label{fig:stability_pressureOM}
\end{center}
\end{figure*}
\begin{figure*}[]
\begin{center}
\includegraphics[scale=0.38]{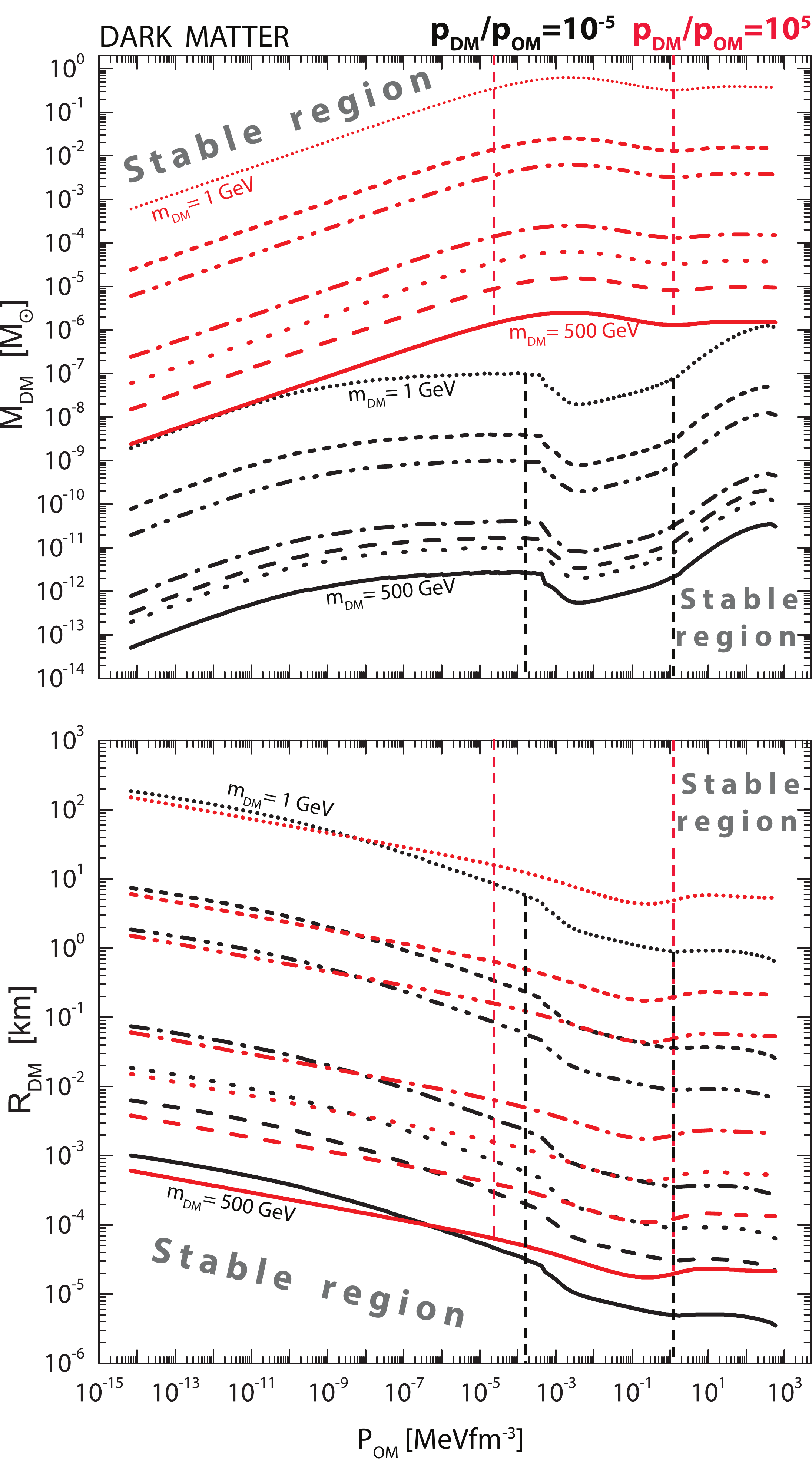}
\includegraphics[scale=0.38]{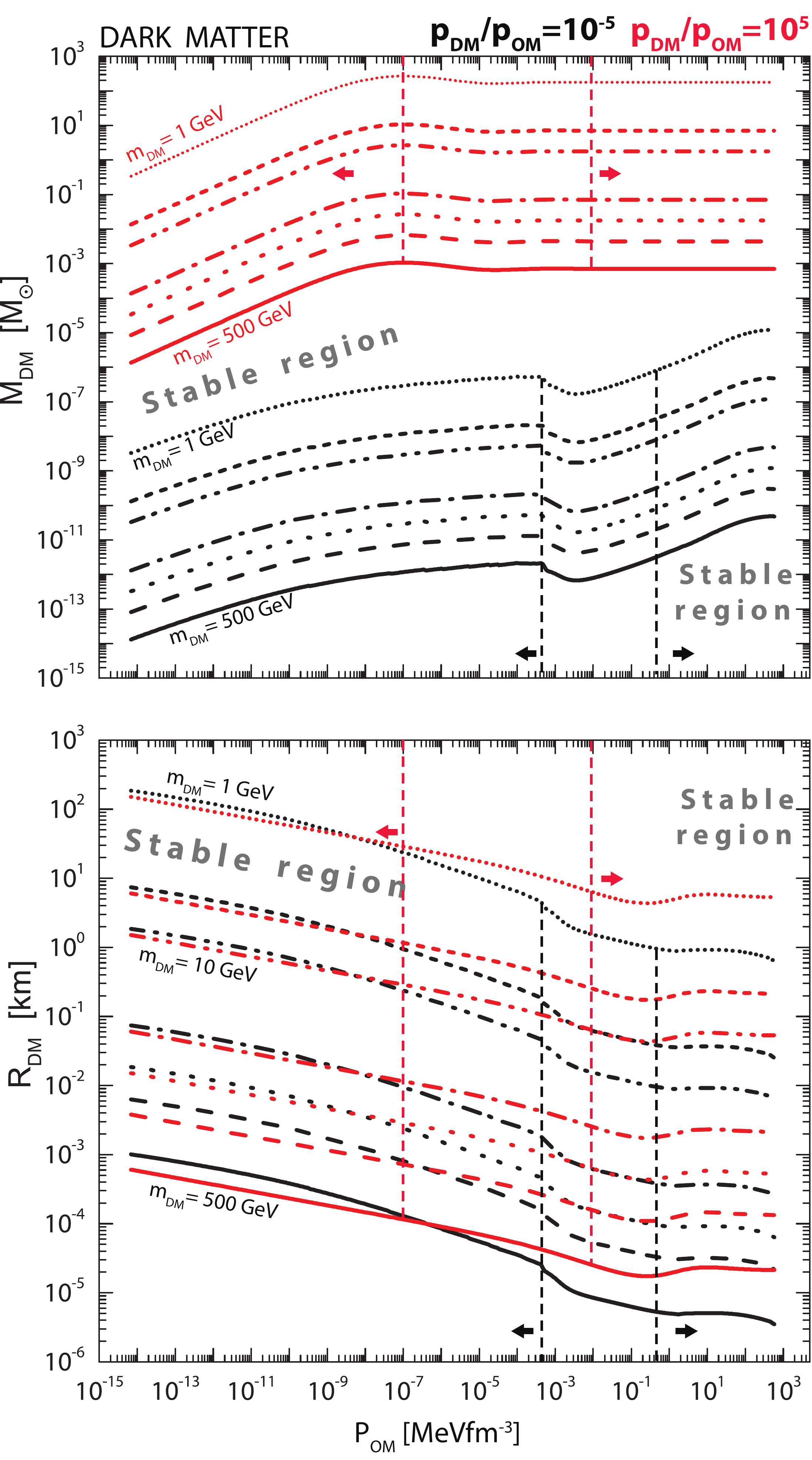}
\caption{
Mass and radius of DM content of the astrophysical objects as a function of the OM central pressure  for the DM interaction strength parameter $y=10^{-1}$ (left panels), and $y=10^{3}$ (right panels).
The vertical black and red dashed lines delimit the stable regions for two {
  sets  given by different $p_{DM}/p_{OM}$,} denoted  by black and red curves. Each curve in each set stands for a DM particle mass going from 1 GeV to 500 GeV (1, 5, 10, 50, 100, 200, 500) from top to bottom. {
  In this figure} black curves correspond to $p_{DM}/p_{OM}=10^{-5}$, whereas red curves correspond to $p_{DM}/p_{OM}=10^{5}$.
}
\label{fig:stability_pressureOMDM_p105}
\end{center}
\end{figure*}
\begin{figure*}[]
\begin{center}
\includegraphics[scale=0.38]{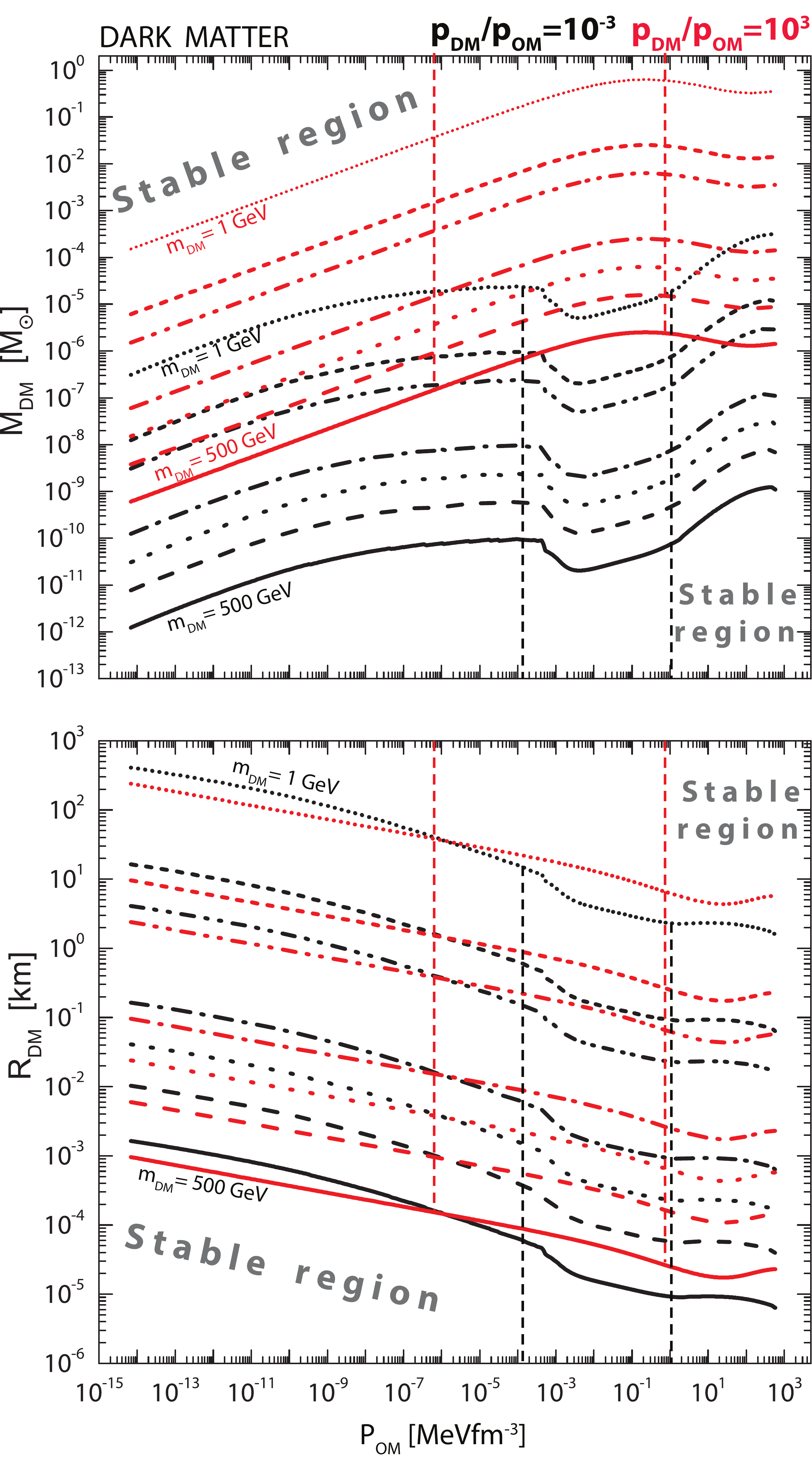}
\includegraphics[scale=0.38]{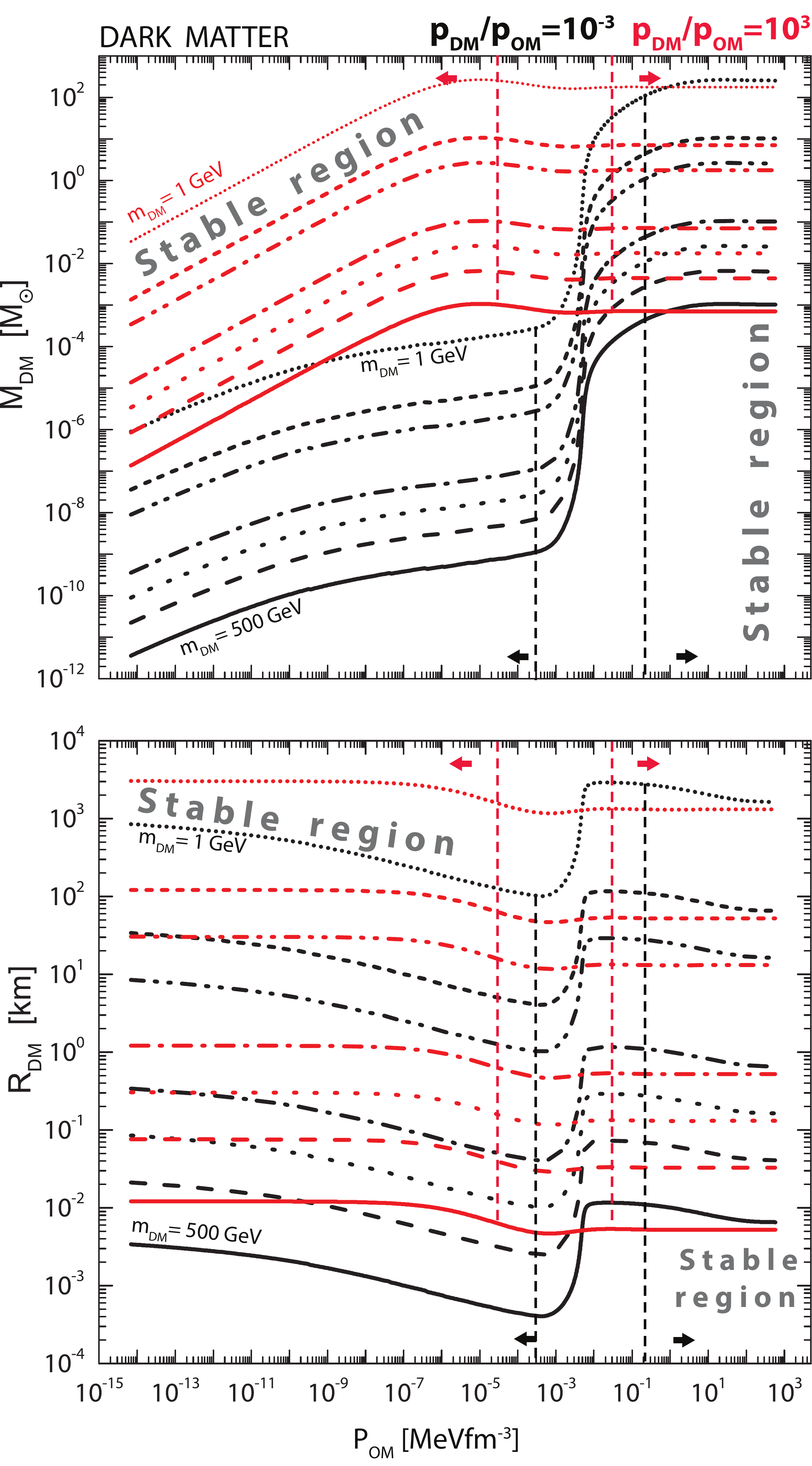}
\caption{
Same as in Fig.\ref{fig:stability_pressureOMDM_p105}, for the dimensionless ratios, $p_{DM}/p_{OM}=10^{-3}$ (black curves) and $10^{3}$ (red curves).
}
\label{fig:stability_pressureOMDM_p103}
\end{center}
\end{figure*}

\begin{figure*}[]
\begin{center}
\includegraphics[scale=0.38]{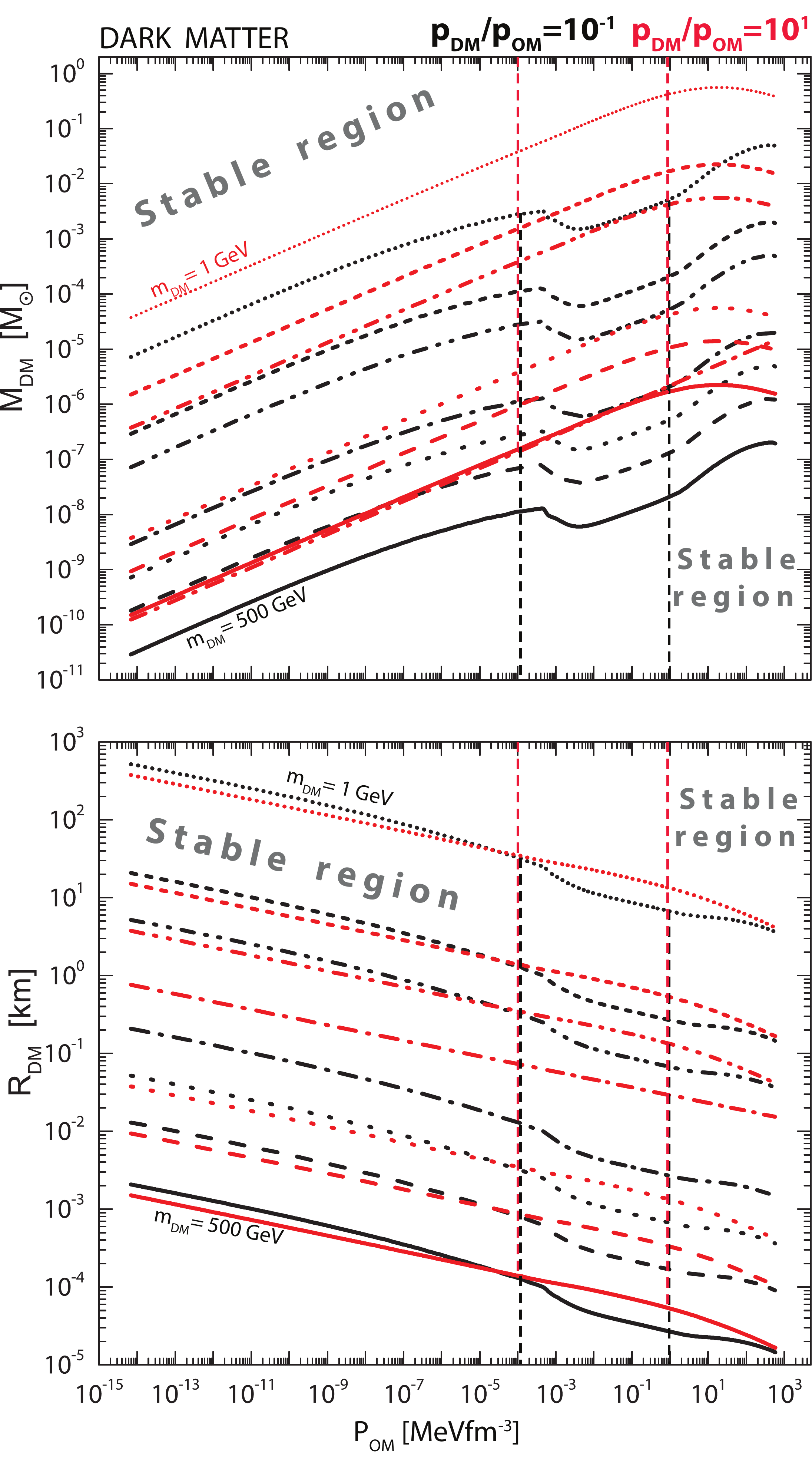}
\includegraphics[scale=0.38]{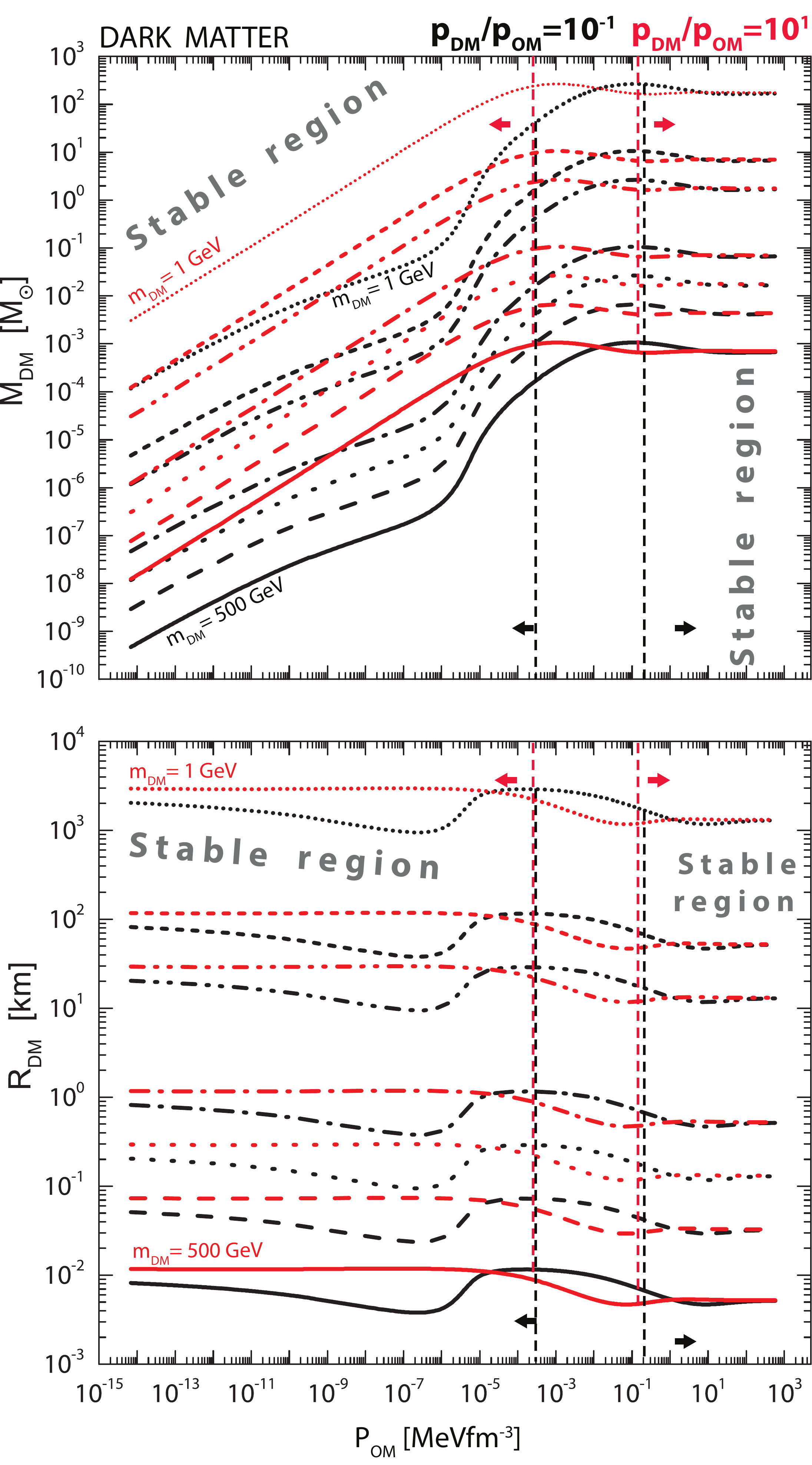}
\caption{
  Same as in Fig.\ref{fig:stability_pressureOMDM_p105}, for the dimensionless ratios, $p_{DM}/p_{OM}=10^{-1}$ (black curves) and $10^{1}$ (red curves).
}
\label{fig:stability_pressureOMDM_p101}
\end{center}
\end{figure*}

{
  For the determination of the COs and their characteristics, one has to perform a stability analysis to determine the possible stable configurations. Our results are given in the next Section, whereas we discuss the stability criteria in 
  Appendix \ref{sec:StabilityDetermination}.}

\section{Results}
\label{sec:Results}


The stable configurations for COs are determined by performing the stability analysis described in 
Appendix \ref{sec:StabilityDetermination} and shown in Figs.~\ref{fig:stability_pressureOM}-\ref{fig:stability_pressureOMDM_p101}
. 
        Whereas Fig.~\ref{fig:stability_pressureOM} shows the stability analysis for OM for different ratios of the dimensionless DM pressure ($p_{DM}$) versus the dimensionless OM pressure ($p_{OM}$), i.e. $p_{DM}/p_{OM}$, each of the Figs.~\ref{fig:stability_pressureOMDM_p105}-\ref{fig:stability_pressureOMDM_p101} shows the analysis for DM for different values of the DM particle mass but for only two $p_{DM}/p_{OM}$ ratios in each figure. In those figures we display the mass (upper panels) and radius (lower panels) of OM (DM) as a function of the central pressure of OM, $P_{\rm OM}$, for weakly interacting matter ($y=10^{-1}$) in the left panels, while the strongly interacting case ($y=10^3$) is shown in the right panels. 

        After performing the stability analysis described in Appendix \ref{sec:StabilityDetermination} for OM only, the stable regions in OM are delimited by vertical lines in both mass and radius plots in Fig.\ref{fig:stability_pressureOM}. As indicated previously, we show the solutions of mass and radius for different $p_{DM}/p_{OM}$. The non-straight feature of the vertical lines in this figure results from joining the different $P_{OM}$ values, for the various $p_{DM}/p_{OM}$ ratios, at which a stable solution turns unstable and vice--versa. We find that the regions in pressure {
          below the first vertical dashed line and above the second vertical dashed one}  give rise to stable mass-radius configurations, for both weakly and strongly interacting DM cases. 


For $P_{\rm OM} \ll 10^{-15}$ MeV fm$^{-3}$, the mass-radius stable configurations on Fig.\ref{fig:stability_pressureOM}, with $M \ll  {\rm M}_{\odot}$ and $M \propto R^{3}$, are of planet-like type \cite{1967hea3.conf..259T}. Subsequently, the mass rises with central pressure while the radius decreases, leading to WD configurations, one of the two stable  $M-R$ branches in compact objects. In this WD configuration and for the case $y=10^{-1}$, the OM central pressure lies below $~10^{-3}$ MeV fm$^{-3}$, while for the case $y=10^{3}$ it remains below $~10^{-5}$ MeV fm$^{-3}$, as clearly displayed in Fig.\ref{fig:stability_pressureOM}. Next, the $M(P_{\rm OM})$ curves present an interval with a series of extrema, with the radius curves changing their slope and giving rise to unstable configurations. Following the stability criteria, the last sharp downturn results in having the last unstable mode turning stable, so that the NS configuration is reached. In this latter region, the OM central pressure is approximately above $10^{-1}$ MeV fm$^{-3}$  for $y=10^{-1}$,  and $~10^{-2}$ MeV fm$^{-3}$ for $y=10^{3}$.  For typical ordinary matter EoS, further increase in central pressure induces the $M-R$ curve to spiral counterclockwise, and more and more modes become unstable. With DM, instead of spiraling at high central pressure, another stable ``twin" or ``third family" branch may arise \cite{Schaeffer:1983, Glendenning:1994sp, Glendenning:1998ag, Schertler:2000xq}.

Once we have restricted the values of $P_{OM}$ for all different $p_{DM}/p_{OM}$ ratios that lead to stable WD and NS configurations with only OM, we analyze the common stability regions for both OM and DM. In this manner, we can apply cuts for the set of the examined DM particle masses (from 1 to 500 GeV). These results are displayed in Figs.~\ref{fig:stability_pressureOMDM_p105} to \ref{fig:stability_pressureOMDM_p101}. In Fig.~\ref{fig:stability_pressureOMDM_p105} we present the results for the mass and radius as function of $P_{OM}$ for  $p_{DM}/p_{OM}=10^{-5}$ and $10^{5}$ for weakly (left panel) and strongly (right panel) interacting DM for DM particle masses ranging from 1 to 500 GeV, whereas in Fig.~\ref{fig:stability_pressureOMDM_p103} we present the cases for $p_{DM}/p_{OM}=10^{-3}$ and $10^{3}$, and $p_{DM}/p_{OM}=10^{-1}$ and $p_{DM}/p_{OM}=10^{1}$ are displayed in Fig.~\ref{fig:stability_pressureOMDM_p101}. 

Each curve in the panels stands for a DM particle mass going from 1 GeV to 500 GeV from top to bottom. Black and red lines correspond to the two sets of $p_{DM}/p_{OM}$, as indicated in these panels. Namely, black curves and black vertical lines correspond to $p_{DM}/p_{OM}=10^{-5}$ (Fig.~\ref{fig:stability_pressureOMDM_p105}), $10^{-3}$(Fig.~\ref{fig:stability_pressureOMDM_p103}), and $10^{-1}$(Fig.~\ref{fig:stability_pressureOMDM_p101}); whereas red curves and red vertical lines correspond to $p_{DM}/p_{OM}=10^{5}$(Fig.~\ref{fig:stability_pressureOMDM_p105}), $10^{3}$(Fig.~\ref{fig:stability_pressureOMDM_p103}), and $10^{1}$(Fig.~\ref{fig:stability_pressureOMDM_p101}). As mentioned previously, the vertical lines are determined by imposing having stable OM and DM simultaneously. The regions {
below the first vertical line and above the second one}, as indicated by the arrows, fulfill the stability criteria for both species. {
{By analyzing the stability region below the lower vertical line, we observe that from $p_{DM}/p_{OM}=10^{-5}$ up to $p_{DM}/p_{OM}=10^{1}$,  the stable region extends up to $P_{OM}\approx10^{-4}$MeV fm$^{-3}$, showing a slight dependence on the ratio $p_{DM}/p_{OM}$, and on the DM interaction strength $y$. By increasing $p_{DM}/p_{OM}$, the size of the stable region shows a strong dependence on $p_{DM}/p_{OM}$ and $y$, and starts well below $p_{OM}\approx10^{-4}$MeV fm$^{-3}$. A similar behaviour is displayed also for the stable region which lies above the second vertical line.}
}
 
\begin{figure*}[]
\begin{center}
\includegraphics[scale=0.32]{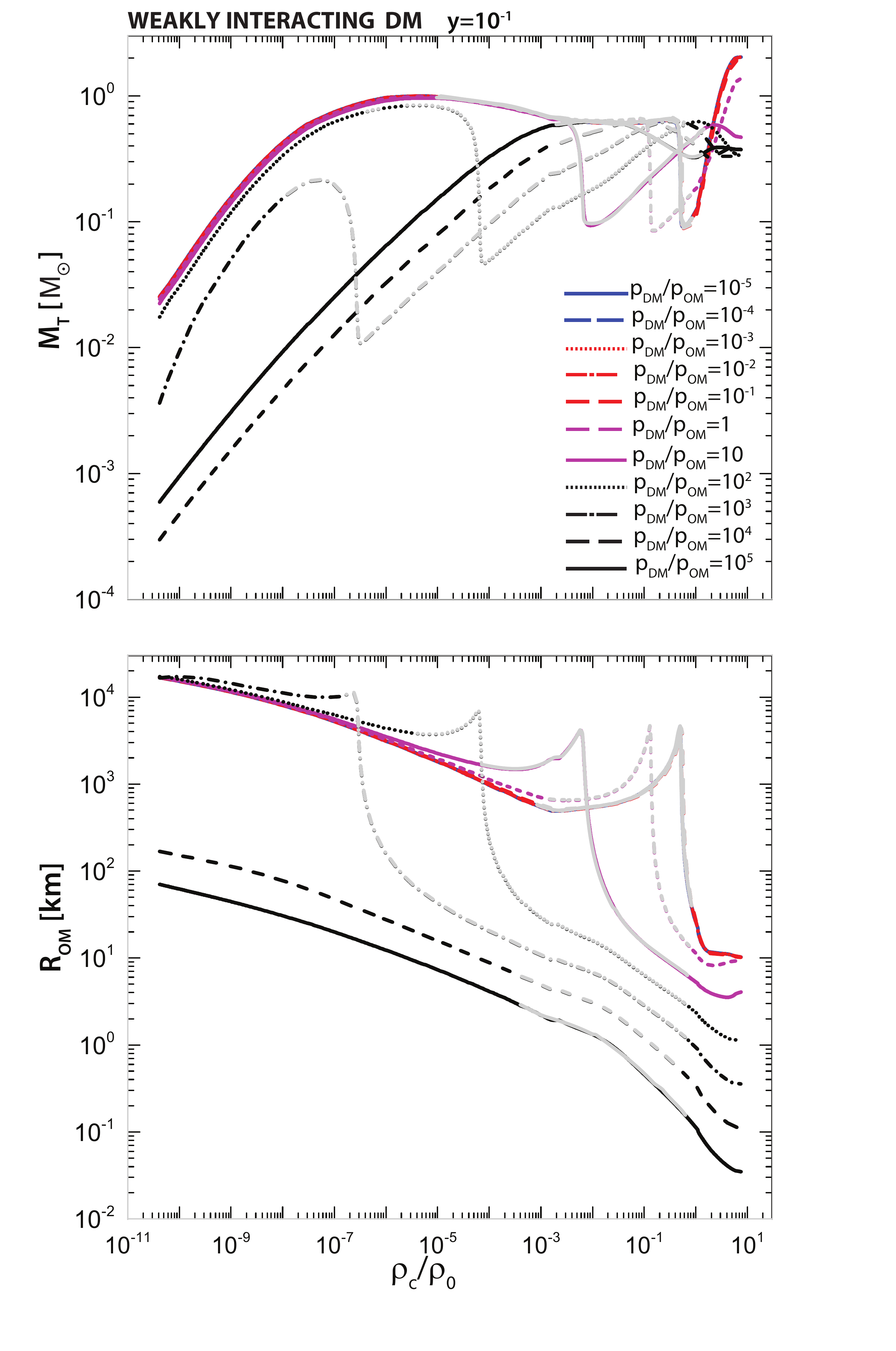}
\includegraphics[scale=0.32]{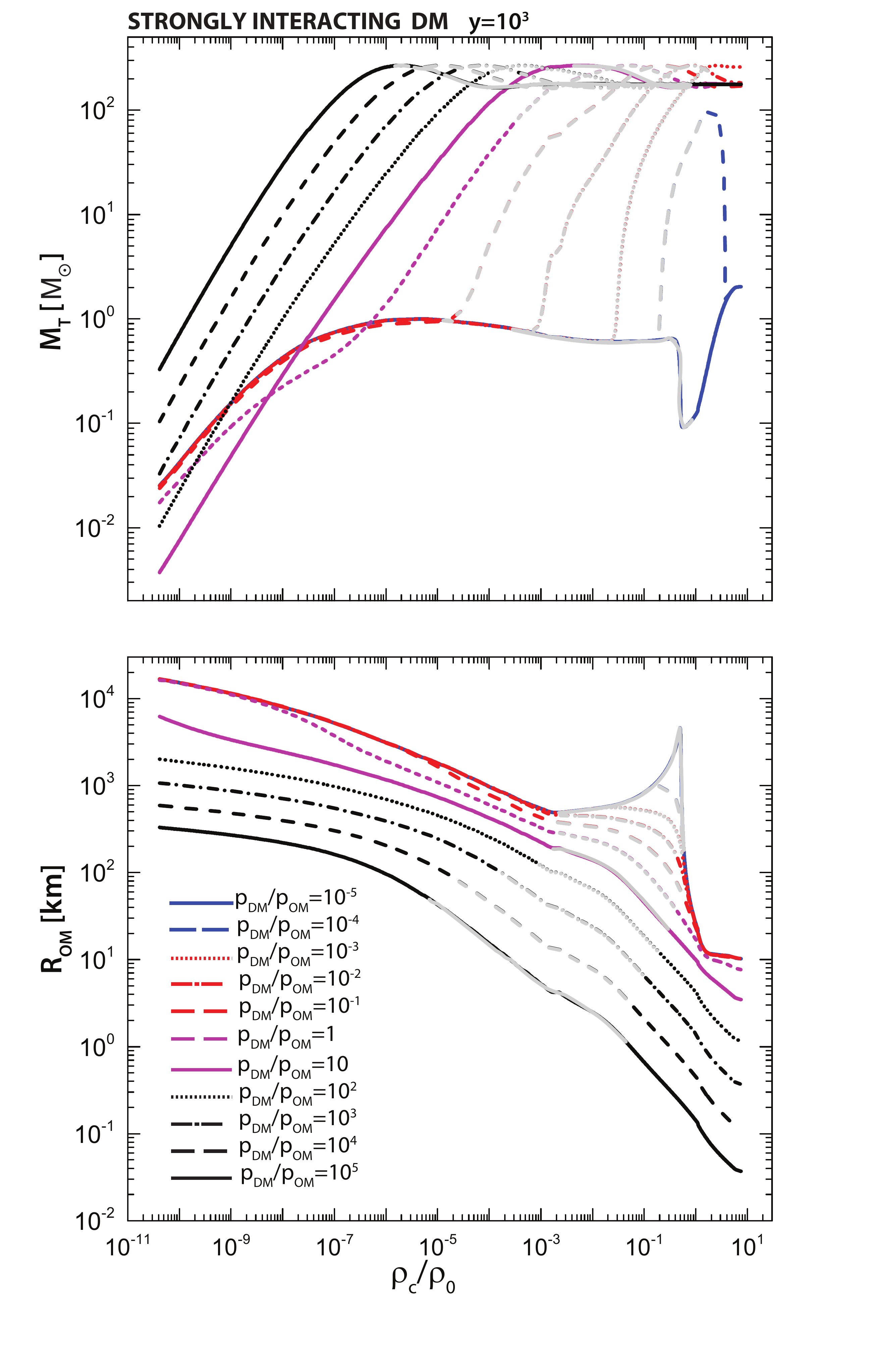}
\caption{Total mass, $M_T$, and radius of OM content, $R_{OM}$, of the astrophysical objects as a function of the central density of OM for the DM interaction strength parameters $y=10^{-1}$ (left panels) and $y=10^{3}$ (right panels). We fix $m_{DM}=1$ GeV. The coloured lines indicate solutions for $M_{T}$ and $R_{OM}$ over OM central densities for a set of dimensionless ratio, $p_{DM}/p_{OM}$, running from $10^{-5}$ to $10^{5}$. 
  These lines turn grey in the unstable regions. Note that for the case $y=10^{-1}$, lines from $p_{DM}/p_{OM} = 10^{-1}$  to  $p_{DM}/p_{OM} = 10^{-5}$ for the $M_{T}$ and $R_{OM}$ distributions lie on top of each other. }
\label{fig:mT_rOM_density}
\end{center}
\end{figure*}

 For completeness, so as to show the range of OM central densities of  our astrophysical objects, we display in Fig.~\ref{fig:mT_rOM_density} the total mass, $M_T$, and the visible radius of OM, $R_{OM}$, of the compact objects as a function of the OM central density, $\rho_{c}/\rho_0$, for  $y=10^{-1}$ (left panels) and $y=10^{3}$ (right panels).  We take  $m_{DM}=1$ GeV. As in Fig.~\ref{fig:stability_pressureOM},  the coloured lines indicate solutions for a set of dimensionless ratio, $p_{DM}/p_{OM}$, running from $10^{-5}$ to $10^{5}$.  These lines turn grey in the unstable regions. 

\begin{figure*}[ht]
\includegraphics[scale=0.35]{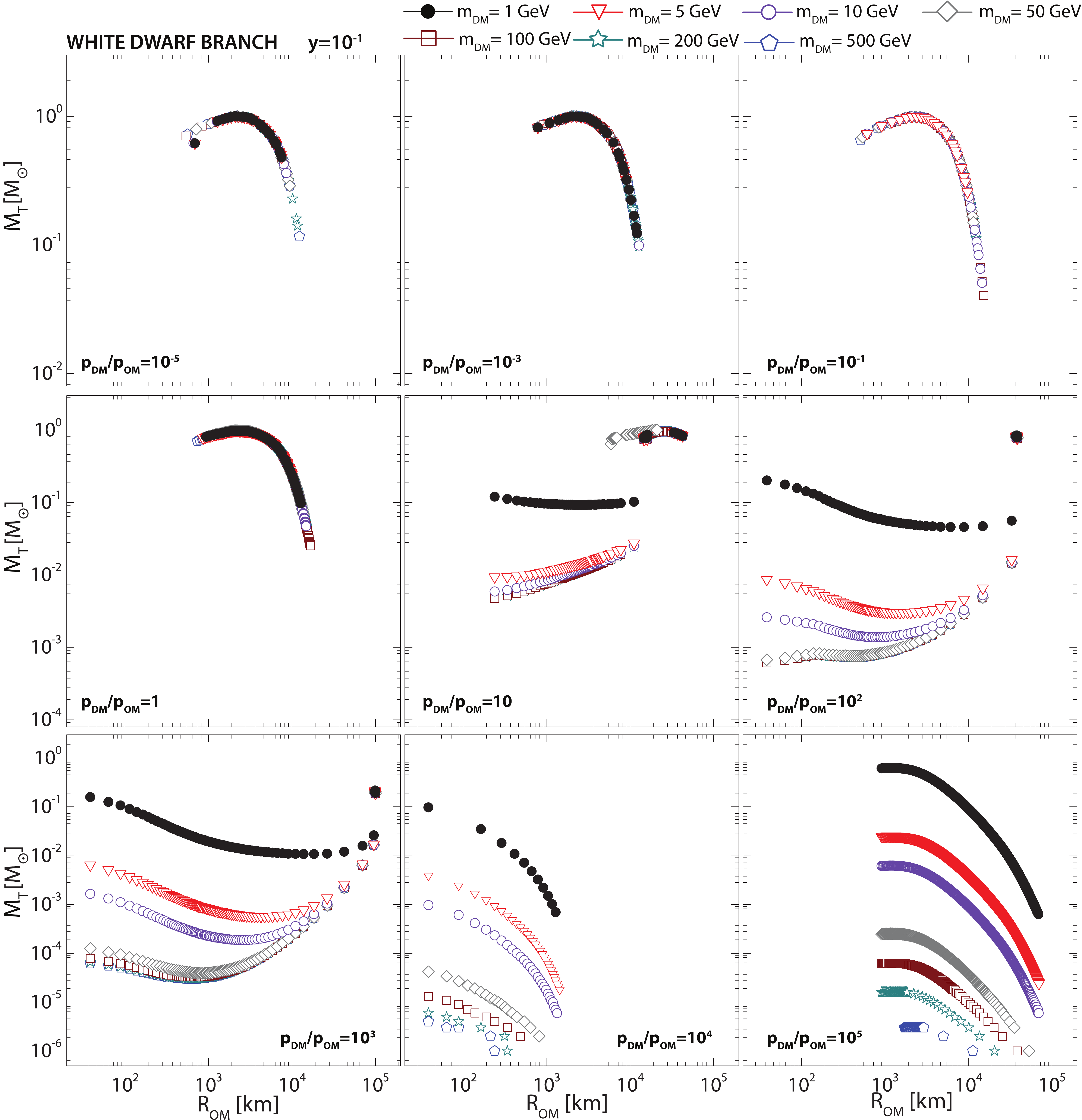}
\caption{
  Mass-radius relations of the equilibrium configuration {
    of  DM-admixed WD branch for } weakly interacting DM, $y=10^{-1}$. Results are shown for DM particle mass $m_{DM}$ ranging from 1 to 500 GeV (1, 5, 10, 50, 100, 200, 500) from top to bottom. 
Each colour corresponds to the DM particle mass, indicated in the legend. Each panel corresponds to a different $p_{DM}/p_{OM}$, indicated in legend. Note that for some values of $p_{DM}/p_{OM}$, some mass curves are not visible because they overlap with others. 
}
\label{fig:radNM_vs_massT_y01_WD}
\end{figure*}

\begin{figure*}[ht]
\includegraphics[scale=0.35]{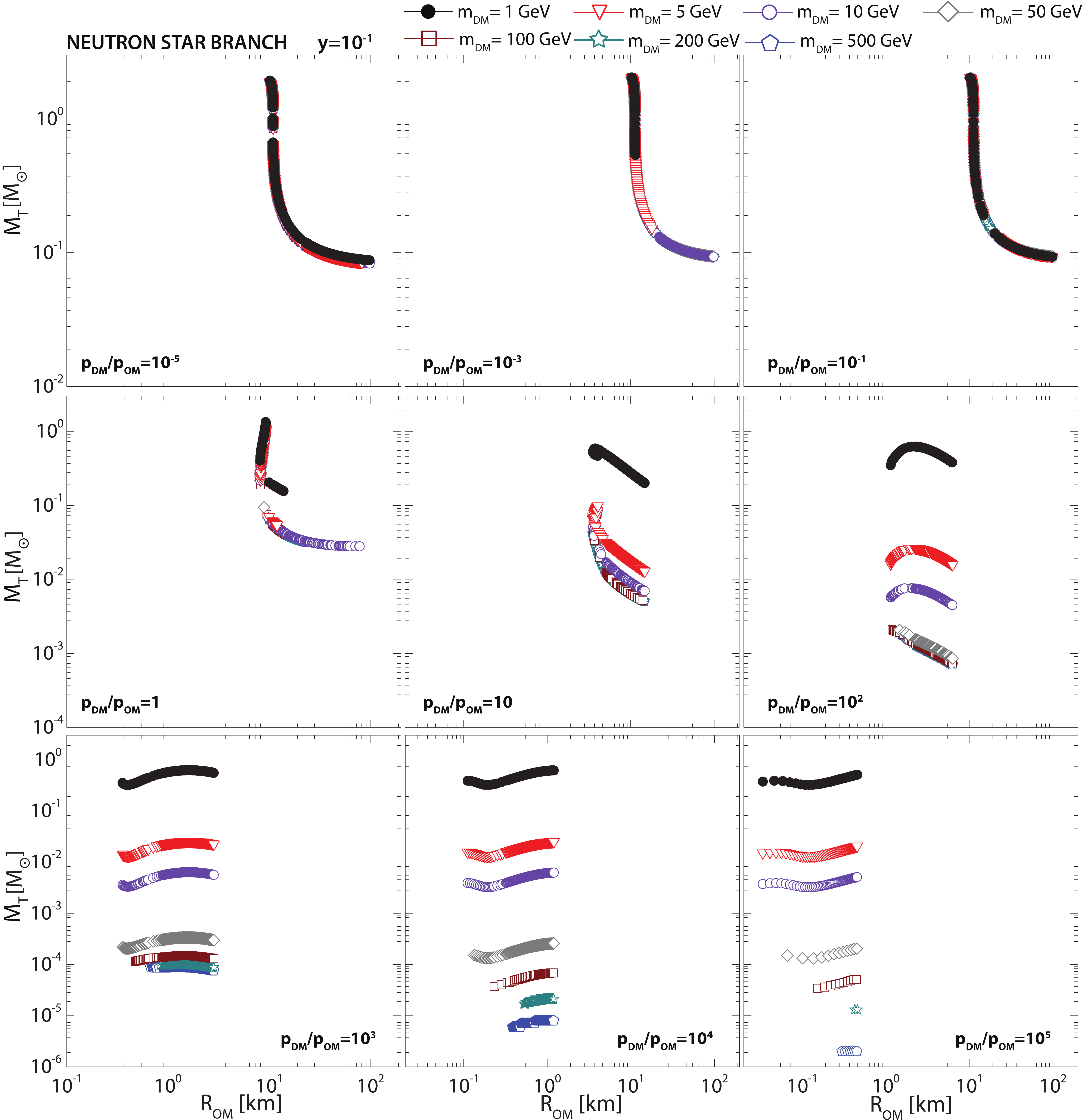}
\caption{
Same as in Fig.~\ref{fig:radNM_vs_massT_y01_WD} for the NS branch. 
}
\label{fig:radNM_vs_massT_y01_NS}
\end{figure*}

\begin{figure*}[ht]
 \includegraphics[scale=0.35]{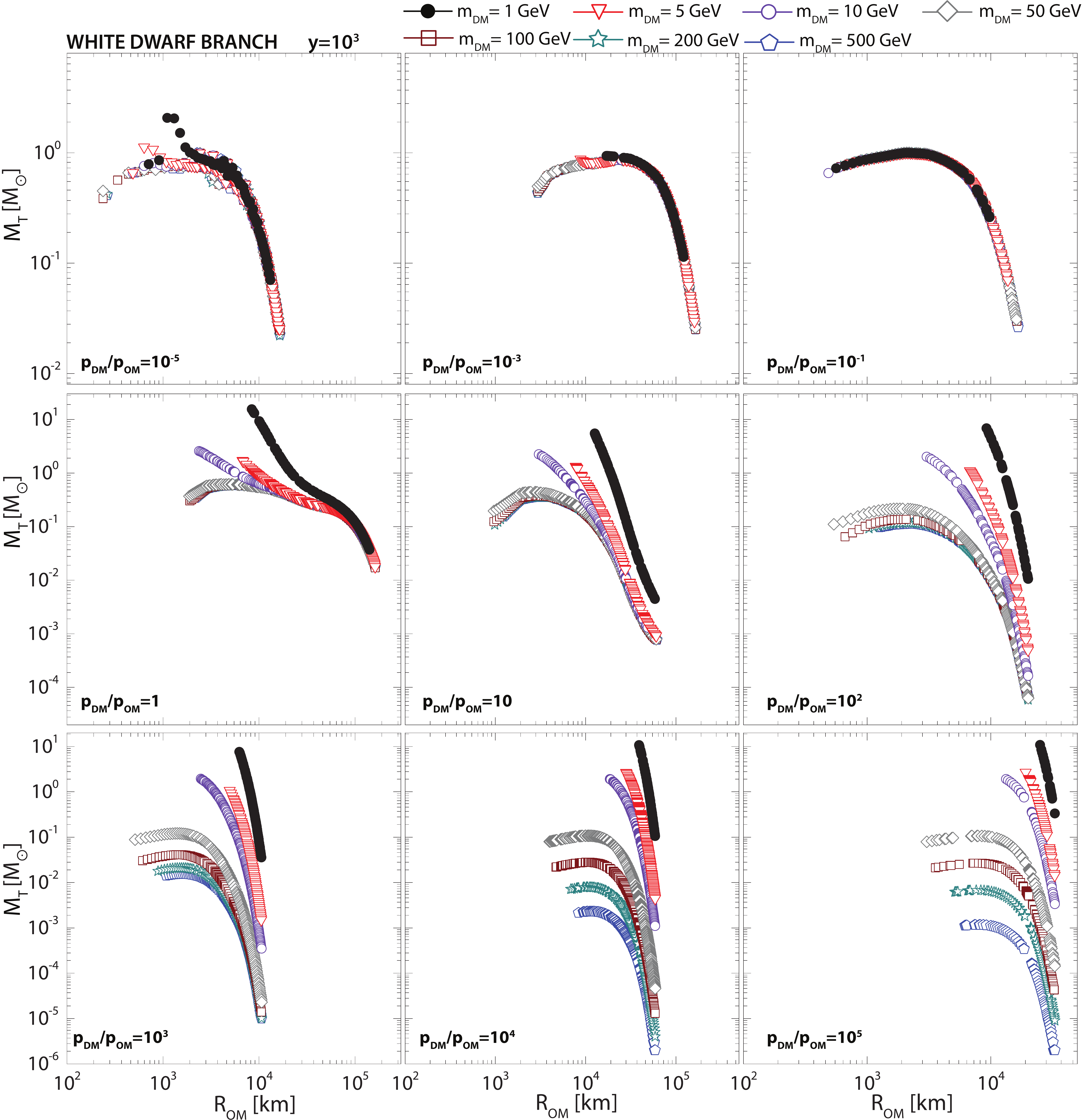}
\caption{
Same as in Fig.~\ref{fig:radNM_vs_massT_y01_WD} for the strongly interacting DM, $y=10^{3}$. 
}
\label{fig:radNM_vs_massT_y103_WD}
\end{figure*}

\begin{figure*}[ht]
\includegraphics[scale=0.35]{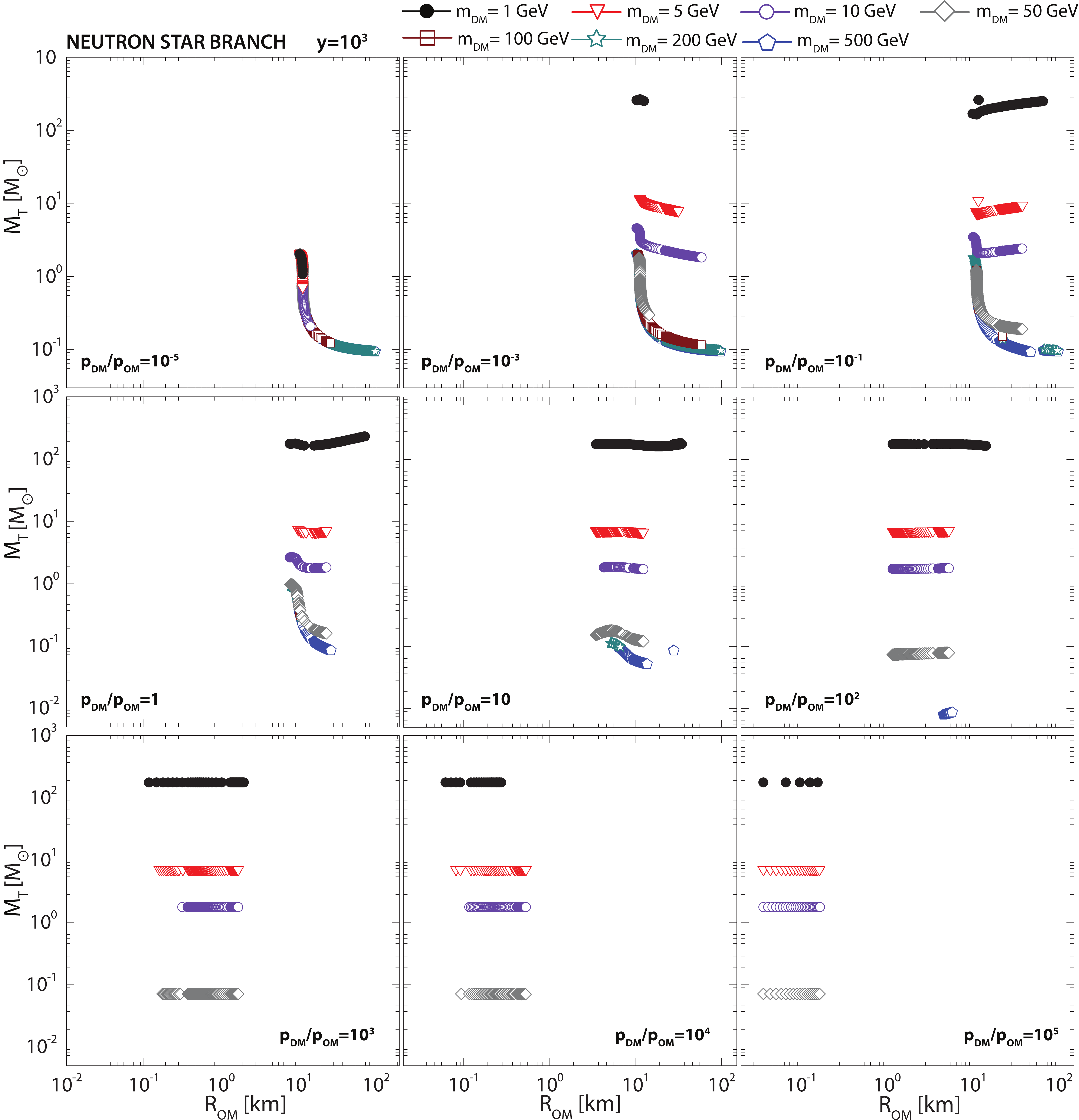}
\caption{
Same as in Fig.~\ref{fig:radNM_vs_massT_y103_WD} for the NS branch.
}
\label{fig:radNM_vs_massT_y103_NS}
\end{figure*}

Once the stability analysis is performed, the features of the COs with DM can be studied from the total mass as a function of the visible radius of OM, that is, $M_{T}$ vs $R_{OM}$, for the NS (high central pressure branch) or WD (low central pressure branch), in a similar manner as done in Ref.~\citep{Tolos2015}. The $M_T-R_{OM}$ relation is presented in Figs.~\ref{fig:radNM_vs_massT_y01_WD}-\ref{fig:radNM_vs_massT_y103_NS} for the range of central pressure ratios ($p_{\rm DM}/p_{\rm OM}=10^{-5}-10^{5}$),  particle mass (1-500 GeV), and  DM interaction strength $y=10^{-1}$, and $y=10^3$.  The cases of weakly interacting DM are shown in Fig.~\ref{fig:radNM_vs_massT_y01_WD} for the WD branch and Fig.~\ref{fig:radNM_vs_massT_y01_NS} for the NS one, while strongly interacting DM yields Fig.~\ref{fig:radNM_vs_massT_y103_WD}  for WD and Fig.~\ref{fig:radNM_vs_massT_y103_NS} for NS. In each figure the panels are ordered left to right and then top to bottom with respect to the sequence of fixed value of pressure ratios, while each panel contains the curves for all the selected DM masses, marked by different colours and symbols.

For ratios of $p_{DM}/p_{OM} < 10^{-1}$ and weakly interacting DM, the two resulting stable $M-R$ configurations are equivalent to the WD and NS branches with only OM  (top panels of Figs.~\ref{fig:radNM_vs_massT_y01_WD} and \ref{fig:radNM_vs_massT_y01_NS}).  While WDs exhibit typical masses of 1 ${\rm M}_{\odot}$ and radii of few thousand km, NSs are characterized by masses of 1-2 ${\rm M}_{\odot}$ and radius of 10 km. With increasing DM central pressure, the NS branch becomes unstable, and the remaining WD branch presents OM densities below the neutron drip line, and unconventional masses and radii (middle and bottom panels). Indeed, we confirm the results of Ref.~\citep{Tolos2015} on COs with Earth-like masses and radii from a few km to a few hundred km in the case of $p_{DM}/p_{OM}=10^{4}$ (bottom middle panel).

As for the strongly interacting case and ratios of $p_{DM}/p_{OM}$ below $10^{-1}$, we observe that the smaller the DM particle mass, the larger the total mass of the CO is, as seen in Ref.~\citep{Sandin2009} for mirror DM. In fact, our results are compatible with the values obtained from Eq.~(47) of Ref.~\citep{Narain:2006kx} for the total mass of an object with DM, that strongly depends on the interaction parameter, and Ref.~\cite{Wang:2018wak}. We again reproduce the COs with Jupiter masses with $10^{-2}-10^{-5} {\rm M}_{\odot}$ masses and few hundred km radii, as found in Ref.~\citep{Tolos2015}.


\begin{figure*}[]
\begin{center}
  \includegraphics[scale=0.7]{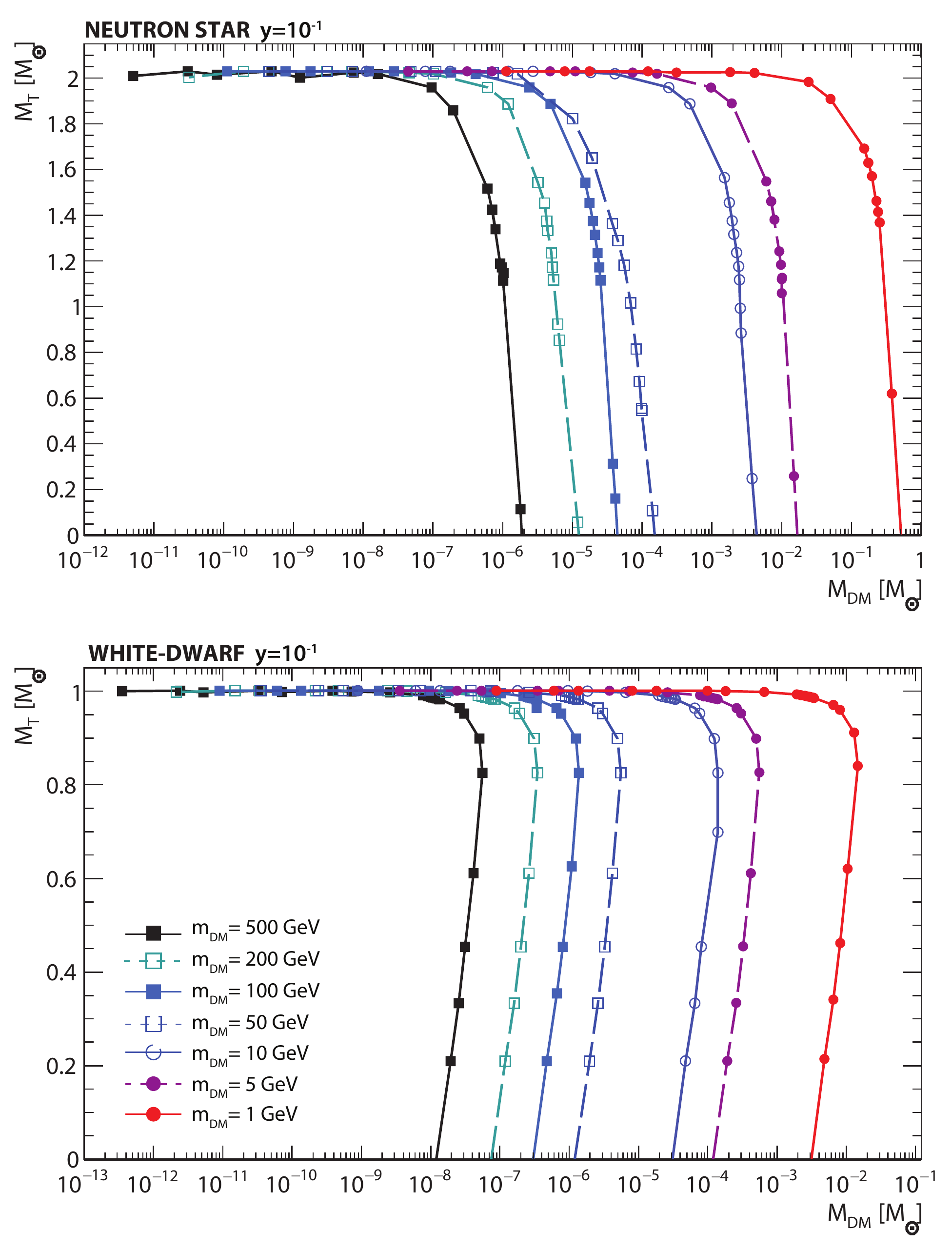}
\caption{
  The {
    total maximum mass} of the CO as a function of the DM mass for weakly interacting DM, $y=10^{-1}$. The coloured curves, built from the extrema at fixed $p_{DM}/p_{OM}$ curves (see text) at a given DM particle mass, show the evolution of the star mass with increasing DM mass content. Each colour corresponds to one DM particle mass, as indicated in legend.
  {\it Top panel}
The neutron-star branches.
  {\it Bottom panel}
  The white-dwarf branches.
}
\label{fig:NSWD_branches_y01}
\end{center}
\end{figure*}

\begin{figure*}[]
\begin{center}
   \includegraphics[scale=0.7]{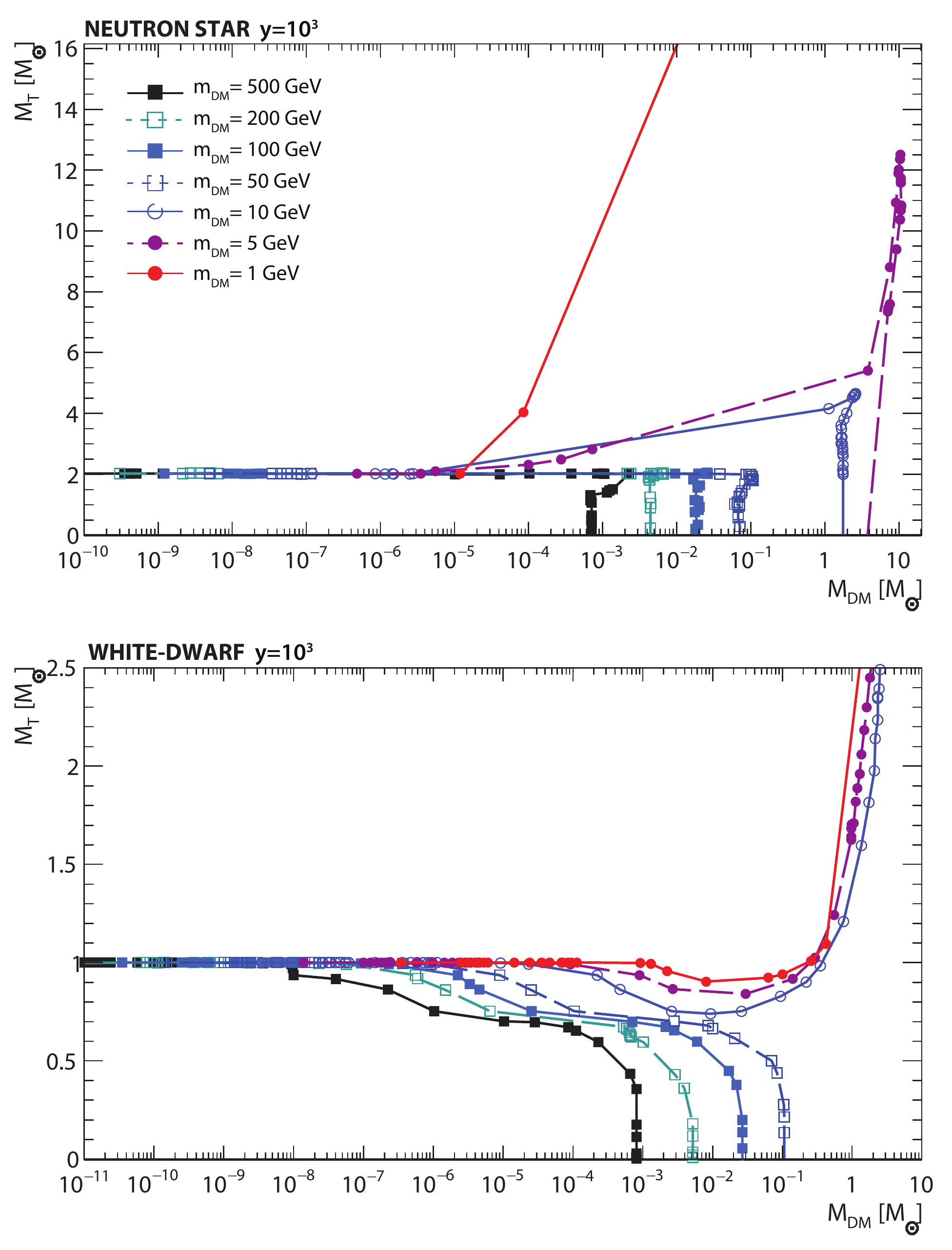}
\caption{
Same as  Fig.~\ref{fig:NSWD_branches_y01} for the strongly interacting DM, $y=10^{3}$. 
Note that lines shift horizontally to the left with increase of DM particle mass, {
  while exhibiting  a growth }with increasing DM content for $m_{DM}=1-10 $ GeV.
}
\label{fig:NSWD_branches_y103}
\end{center}
\end{figure*}

Finally, we show the maximum total masses of the COs in the NS and the WD branches as function of the DM mass content, {
  $M^{\rm max}_{T}(M_{DM})$}, for weakly interacting DM (Fig.~\ref{fig:NSWD_branches_y01})  and strongly interacting DM (Fig.~\ref{fig:NSWD_branches_y103}). The upper panels in these figures show the NS branch, whereas the low panels display the WD one. The different curves in each panel are given for a fixed value of the DM mass particle.  The maximum mass values of the CO are obtained from the maximum masses of all possible stable NS and WD configurations given by a fixed $p_{DM}/p_{OM}$ ratio, but varying the DM particle masses (1-500 GeV) and for weakly and strongly-interacting DM matter, as can be extracted from  Figs.~\ref{fig:radNM_vs_massT_y01_WD}-\ref{fig:radNM_vs_massT_y103_NS}.



As seen in Ref.~\citep{Tolos2015}, we observe that, for the weakly interacting DM case, the reduction of the total mass in the WD branch from the nominal value of 1$M_{\odot}$ takes place for lower DM mass content than for the case of 2$M_{\odot}$ in NSs. Therefore, WDs sustain less DM than the most massive NSs. This is also the case for strongly interacting DM for DM particles with masses above 50 GeV. On the other hand, for DM particle masses of 10 GeV and less, we obtain an increase in the total mass of CO when DM content exceeds $\sim 10^{-5} M_{\odot}$ for the NS branch and $\sim 10^{-1} M_{\odot}$ for WDs. This might exclude strongly self-interacting DM of masses 1-10 GeV in the interior of COs if formation mechanisms of COs allow for these DM mass contents. The possible  formation of these COs and their DM content is discussed in the following section.

\section{Matter in the COs: accumulation and environment.} 
\label{sec:Matter in the COs}


In order COs having terrestrial or Jupiter-like masses, they should acquire from the environment a precise quantity of DM. As an example, for an object like Jupiter having a mass $\simeq 10^{-3} {\rm M}_{\odot}$, the content of DM must be in the range of $10^{-1}-10^{-5} {\rm M}_{\odot}$, as found in Ref.~\citep{Tolos2015} from the solution of the TOV. One question that can naturally arise is if, in nature, there are processes that allow the accretion of that quantity of DM by a CO.

To have an idea of the quoted quantity, we have to consider the DM acquired during the different phases of the CO formation. In the case of a Earth-like, or Jupiter-like CO, there are two phases of accretion: a) accretion during the collapse phase; b) accretion of DM after collapse due to the capture in the CO by interaction with the CO's nuclei. In the case of NSs or WDs, one must take account of a) the DM acquired during the collapse phase, b) the change of DM inside the star during the star evolution till the supernova explosion phase, c) the DM acquired by the NS. 

In order to have a good estimate of the DM in NSs and WDs, one should perform simulations similar to those of \citep{Yang2011}, but for the inner part of the NSs and WDs, not only the distribution of DM external (mini-halo) to the stars. Simplified calculation consider just the DM capture during the NS, WD phase \citep{Kouvaris2008,Kouvaris2013}, or estimate the accretion by the CO progenitor, and the CO phase \citep{Kouvaris2010}. 
One can obtain an estimate of the DM trapped in the star by using Eq.~(4) in \citep{Kouvaris2011} for the capture rate
\begin{eqnarray}
F &=& 1.1 \times 10^{27} s^{-1}  \left( \frac{\rho_{dm}}{ 0.3 {{\rm{GeV/cm^{3}}}} } \right) \left(\frac{220 {{\rm{km/s}}}}{v}\right) \left(\frac{{{\rm{TeV}}}}{m} \right)\nonumber \\
&& \left(\frac{M}{M_{\odot}}\right) \left(\frac{R}{R_{\odot}}\right)  \left(1-e^\frac{-3E_0}{v^2}\right) f,
\label{Kuv}
\end{eqnarray}
where $\rho_{dm}$ is the local dark matter density, $v$ the average WIMP velocity, $f$ the probability that in the star one has at least one WIMP-proton scattering and {$E_0$ the maximum energy of the WIMP per WIMP mass leading to capture} (see \citep{Kouvaris2011}). In the case of a typical NS (WD), Eq.\eqref{Kuv} gives $\simeq 10^{-14} (10^{-11}) {\rm M}_{\odot}$, for the DM trapped in a NS (WD), and values even smaller in the case of the planet-like COs.  The capture rate has been estimated by several other authors \citep{Kouvaris2013,Zhong2012,Zheng2016,Guver2014} and the results are more or less in agreement with that of \citep{Kouvaris2011}. A comparison of the DM contained in the planet-like COs, and the DM that can be trapped into them by accretion shows that this  mechanism cannot explain, at first glance, their existence (see however the following), in the case DM is uniformly distributed in the halo. {
  The previous discussion} shows a discrepancy between what the TOV equation allows, and the quantity of DM that the accretion mechanism can trap in the star.

As many studies \citep{Berezinsky2003,Ricotti2009,Scott2009,Berezinsky2013,Bringmann2012,Berezinsky2014}
pointed out, the DM distribution into an halo is not homogeneous, and super-dense dark matter clumps (SDMC), i.e. bounded DM objects virialized at the radiation dominated era, and ultra compact mini-haloes (UCMH) formed from  secondary accretion on SDMCs \citep{Berezinsky2013}, are present in the halo. 

According to those studies, a SDMC of $\simeq 100 {\rm M}_{\odot}$ can have a density  
$\simeq 2 \times 10^6$ larger than the local DM density, and larger for smaller masses. A NS located in such a SDMC would trap through the accretion process a DM mass of the order of $10^{-7} {\rm M}_{\odot}$, $10^{-6} {\rm M}_{\odot}$ for a SMDC of 0.1 ${\rm M}_{\odot}$ \citep{Berezinsky2013,Berezinsky2014}. 

The maximum density in the center of clumps can be estimated by means of the annihilation criterion \citep{Berezinsky2013,Berezinsky2014} and is $\simeq 10^{10}$ larger that the local DM density, implying that a NS would acquire a DM mass equal to $\simeq 7.5 \times 10^{-4} {\rm M}_{\odot}$.  
%
%
~\\

In the previous scenario, we considered COs planet-like, either NSs or WDs, which trapped DM from the environment by means of the accretion mechanism. However, another possibility, actually the more natural, is that  mini-halo forms and a planet-like CO is born by the following collapse of baryons on the potential well of the mini-halo. The last correspond to the phase a. (accretion during the collapse phase) previously mentioned.  After formation, the planet could continue to acquire DM by accretion from the environment (phase b). We will study this aspect in a future work.

Finally, even if DM was distributed in a homogeneous fashion in the Galaxy, its density increases going toward the center of the Galaxy, similarly to what happens in a mini-halo. The DM density profile of our Galaxy is not known. In particular it is not known if the profile is cored, as in several dwarf spiral galaxies, or cuspy. N-body simulations predict cuspy profiles parameterized by an Einasto profile: 
\begin{equation}
\rho=\rho_{{-2}}{{\rm e}^{-2\,{\frac {1}{\alpha} \left[  \left[ {\frac {r}{r_{
{-2}}}} \right] ^{\alpha}-1 \right] }}},
\end{equation}
where $\alpha$ is related to the curvature degree of the profile, $r_{-2}$ is the distance at which $\frac{d \ln{\rho}}{d \ln{r}}=-2$, and $\rho_{{-2}}$ is the density at $r_{{-2}}$. 
In the case of the more cuspy profile  (see \citep{deLavallaz2010}), the DM density at $10^{-5}$ pc is $4 \times 10^9$ GeV/cm$^3$, a factor $\simeq 10^{10}$ larger than in the Sun neighborhood. This means that while a NS located at the Sun neighborhood will accrete $\simeq 10^{-14} {\rm M}_{\odot}$, at $10^{-5} (0.1)$ pc will accrete $10^{-4} (10^{-7}) {\rm M}_{\odot}$. This implies that even in a homogeneous halo, planet-like CO objects can form at about $0.1$ pc to the center of the Galaxy.

The previous discussion has some consequences on the NSs and WDs structure, and generates a 
relationship between the COs masses and the distance from galactic center, which will be discussed in a forthcoming paper.


\section{Conclusions and discussions}
\label{sec:Conclusions}


In this paper, we have studied how DM, non-self annihilating, self-interacting dark matter, admixed with ordinary matter in COs 
changes their inner structure, and discussed the formation of planet-like COs. We consider DM particle masses from 1 to 500 GeV, while taking into account both weakly and strongly interacting DM.

The total mass of the COs depends on the DM particle mass as well as the quantity of DM in its interior. In the strong interacting case, some combinations of DM particle mass and DM mass content lead to COs much heavier than  the known limits given by NSs and WDs. This puts constraints in the parameter space $y$, $M_D$, $p_{\rm DM}/p_{\rm OM}$, and $R_{\rm OM}$. {
  Also, constraints on the mass trapped in the pulsar come from the observations of pulsars with masses $\simeq 2 M_{\odot}$. } At the same time, it is  possible to explain  the existence of NSs with masses $\simeq 2 M_{\odot}$ through the increase of the total COs mass with decrease of particle mass. 


{
  
{ Following the discussion in Section \ref{sec:Matter in the COs}, on the inhomogeneity of DM distribution in the halo, and minihaloes of the Milky Way, in a next paper we will discuss how the inhomogeneity affects the COs structure. As we indicate in Section \ref{sec:Matter in the COs},   we know that DM is not uniformly distributed inside the galactic haloes for two reasons: a) the presence of clumps (mini-haloes) randomly located inside the galactic halo, b) the increase of DM density going towards the center of the galactic halo. The difference from outskirts to the halo, or mini-halo, center is usually of $10^9$, $10^{10}$. This implies the accumulation mechanism can trap much more mass close to the halo center, giving rise to different COs structures. }


\section*{Acknowledgments}

{
M.D. work was supported by the Chinese Academy of Sciences Presidents International Fellowship Initiative Under Grant No. 2016PM043 and by the Polish National Science Centre (NCN) grant 2016/23/B/ST2/00692. MLeD acknowledges the financial  support by Lanzhou University starting fund.
L.T. acknowledges support from the FPA2016-81114-P Grant from Ministerio de Ciencia, Innovacion y Universidades, Heisenberg Programme of the Deutsche Forschungsgemeinschaft under the Project Nr. 383452331 and PHAROS COST Action CA16214. }

\appendix
\section{Appendix: Stability Determination}
\label{sec:StabilityDetermination}


{

  The stability of TOV solutions is commonly determined using one of two methods. {
    The first is the Bardeen, Thorne, and Meltzer (BTM) method \cite{Bardeen:1966aj}, based} on counting the mass-radius ($M-R$) relation extrema, which curves are generated by varying the central pressure $P_{c}$. 
Each curve point is a stationary configuration, but only those with stable radial modes are stably stationary. The simple rules proposed by Ref.~\cite{Bardeen:1966aj} are: at each extremum, with increasing central pressure, one mode changes from i) stable to unstable where the $M(R)$ curve rotates counter-clockwise; ii) unstable to stable where the $M(R)$ curve rotates clockwise.
The other method consists in solving the relevant Sturm-Liouville equation to explicitly obtain the radial oscillation eigenmodes, following \cite{Shapiro:1983du}. This is used by \citep{Tolos2015, Alford:2017vca} and by us in what follows. 

The time-dependent displacement describes the radial oscillations:
\begin{equation}
\delta r_{n}(r,t)=\frac{e^{\nu(r)}}{r^{2}} u_{n}(r)e^{i\omega_{n}t},
\label{eq:radial_osc}
\end{equation}
where $\nu(r)$ comes from the double radial metric definition, $n$ marks the mode spectrum index, and $u_{n}(r)$ is a solution with eigenvalue $w_{n}^{2}$ to the Sturm-Liouville problem
\begin{equation}
\frac{d}{dr} \left( \Gamma P \frac{1}{r^{2}}\frac{d}{dr} ( r^{2}\xi ) \right) - \frac{4}{r}\frac{dP}{dr}\xi + \omega^{2}\rho_{c}\xi =0,
\label{eq:SL_eigenvalue_probl}
\end{equation}
where $\Gamma=\frac{d {\rm{log}} P(r)/dr}{d {\rm{log}} \rho(r)/dr}$ gives the adiabatic index governing the pressure-density equilibrium relation. 
More precisely, it yields its pressure-weighted average, i.e. the fractional change in pressure per fractional change in comoving volume, at constant entropy and composition. The quantity $\xi$ gives the radial component of the perturbations $\vec{\xi}(\vec{x},t)$, $\rho_{c}$ is the central mass density, and $P$ is the electrostatic pressure.

The boundary conditions for Eq.~\eqref{eq:SL_eigenvalue_probl} are 
$\xi=0$  at  $r=0,$ and $\xi$ finite  at $r=R$, 
where $R$ is the surface of the star.

The eigenvalues of Eq.~\eqref{eq:SL_eigenvalue_probl} are given by
\begin{equation}
\omega^{2} = \frac{\int_{0}^{R}\{ \Gamma P \frac{1}{r^{2}}   \left[  \frac{d}{dr} (r^{2}\xi)^{2} + 4r\xi^{2} \frac{dP}{dr} \right] \}dr}{\int_{0}^{R}\rho_{c} \xi^{2} r^{2} dr},
\label{eq:SL_w0}
\end{equation}
where we have integrated by parts, using the boundary conditions $\xi=0$ at $r=0$ and $\Delta P=0$ at $r=R$. The physical interpretation of some terms in Eq.\eqref{eq:SL_w0} follows. Since we always have $\Gamma\ge0$, the first term is stabilizing. It arises from the electric field ``compression" or the electrostatic pressure. It can be equivalently interpreted as due to the acoustic modes. The second term reflects gravity destabilizing effect. 

The Sturm-Liouville eigenvalue problem, Eq.~\eqref{eq:SL_eigenvalue_probl}, produces a discrete set of solutions eigenfunctions $u_{n}(r)$, for eigenvalues $\omega^{2}$ from Eq.~\eqref{eq:SL_w0}, the squared frequencies of the oscillation modes. Those eigenvalues form a real lower-bounded infinite sequence $\omega_n^2 < \omega_{n+1}^2$, for $n=0,1,2...$.  
Any mode $n$ is stable and oscillatory if $\omega_{n}^{2} \textgreater 0$, so the frequency is real. 
However, for $\omega_{n}^{2} \textless 0$, the purely imaginary frequency leads to an unstable mode which exponentially grows or decays.

The overall stability of the star is sufficiently determined by the fundamental radial mode, $\omega_{0}^{2}$. Indeed, if $\omega_{0}^{2} \textgreater 0$, then all $\omega_{n}^{2} \textgreater 0$ and the star is stable. For $\omega_{0}^{2} \textless 0$, there is (at least) one unstable mode and the star is unstable. The sign of $\omega_{0}^{2}$ thus sufficiently ascertain the overall stability. It derives from the analysis of the star mass versus mass density or radius \cite{Shapiro:1983du}.  The typical behavior of the lowest eigenvalues, $\omega^{2}_{0}$ and $\omega^{2}_{1}$ can be found in Ref.~\cite{Alford:2017vca}.



Sturm-Liouville properties of the perturbation equation yield general arguments showing that the mode stability changing extremum occurs for even modes ($n=0,2,...$) if $dR/d\rho_{c} \textless 0$  and for odd modes ($n=1,3,...$) if $dR/d\rho_{c} \textgreater 0$. {
  Using this method and starting from low-mass densities where all modes are positive, the stability analysis is obtained for higher-mass densities studying the change of the sign of the different modes  while keeping their hierarchy. }
 
}


\bibliographystyle{apsrev4-1}
\bibliography{old_MasterBib2}
\end{document}